%% file: main.tex
\documentclass[twocolumn]{aastex63}
\shorttitle{Black Hole Accretion and Jet Timescales of the Young $z\sim6$ Quasar P352-15}
\shortauthors{Rojas-Ruiz et al.}

\usepackage{float}
\usepackage{natbib}
\usepackage{hyperref}
\usepackage[nointegrals]{wasysym}
\usepackage{ragged2e}
\usepackage{graphicx}
\usepackage{units}
\usepackage{amsmath}
\input{definitions}

\begin{document}

\title{First Measurements of Black Hole Accretion and Radio-jet Timescales in a Young Quasar at the Edge of Reionization}

\correspondingauthor{Sof\'ia Rojas-Ruiz}
\email{rojas@astro.ucla.edu}

\author[0000-0003-2349-9310]{Sof\'ia Rojas-Ruiz}
\affiliation{Max-Planck-Institut f\"{u}r Astronomie, K\"{o}nigstuhl 17, D-69117, Heidelberg, Germany}
\affiliation{Department of Physics and Astronomy, University of California, Los Angeles, 430 Portola Plaza, Los Angeles, CA 90095, USA}

\author[0000-0003-3168-5922]{Emmanuel Momjian}
\affiliation{National Radio Astronomy Observatory, P.O. Box O, Socorro, NM 87801, USA}

\author[0000-0003-0821-3644]{Frederick B. Davies}
\affiliation{Max-Planck-Institut f\"{u}r Astronomie, K\"{o}nigstuhl 17, D-69117, Heidelberg, Germany}

\author[0000-0002-2931-7824]{Eduardo Ba\~nados}
\affiliation{Max-Planck-Institut f\"{u}r Astronomie, K\"{o}nigstuhl 17, D-69117, Heidelberg, Germany}

\author[0000-0003-2895-6218]{Anna-Christina Eilers}
\affiliation{MIT Kavli Institute for Astrophysics and Space Research, 77 Massachusetts Ave., Cambridge, MA 02139, USA}

\author[0000-0001-8582-7012]{Sarah B. Bosman}
\affiliation{Institute for Theoretical Physics, Heidelberg University, Philosophenweg 12, D-69120, Heidelberg, Germany}

\author[0000-0001-5424-0059]{Bhargav Vaidya}
\affiliation{Department of Astronomy, Astrophysics and Space Engineering, Indian Institute of Technology Indore, 453552, India}

\author[0000-0001-6647-3861]{Chris Carilli}
\affiliation{National Radio Astronomy Observatory, P.O. Box O, Socorro, NM 87801, USA}

\author[0000-0002-5941-5214]{Chiara Mazzucchelli}
\affiliation{Instituto de Estudios Astrof\'{\i}sicos, Facultad de Ingenier\'{\i}a y Ciencias, Universidad Diego Portales, Avenida Ejercito Libertador 441, Santiago, Chile.}

\author[0000-0002-7898-7664]{Thomas Connor}
\affiliation{Center for Astrophysics $\vert$\ Harvard\ \&\ Smithsonian, 60 Garden St., Cambridge, MA 02138, USA}

\author[0000-0002-7220-397X]{Yana Khusanova}
\affiliation{Max-Planck-Institut f\"{u}r Astronomie, K\"{o}nigstuhl 17, D-69117, Heidelberg, Germany}

\begin{abstract}
We present the first study dedicated to measuring the timescales for black hole accretion and jet launch in a quasar at the edge of Reionization, PSO J352.4034--15.3373 at $z\!=\!5.832 \pm 0.001$. 
Previous work presented evidence of the strong radio synchrotron emission from the jet affecting the host galaxy dust-dominated continuum emission at $\nu_{\rm rest}=683$\,GHz ($\nu_{\rm obs}=100$\,GHz), implying a break in the synchrotron spectrum. In this work, we present quasi-simultaneous observations at 1.5\, GHz -- 42\,GHz with the Karl G. Jansky Very Large Array\,(VLA), and derive a frequency break at $\nu^{\rm break}_{\rm rest} = 196.46$\,GHz ($\nu^{\rm break}_{\rm obs} = 28.76$\,GHz). Modeling these observations, we calculate the jet spectral aging from the cooling of electrons to be $t_{\mathrm{spec}}\sim 580$\,yr. From this measurement, we approximate the dynamical age $t_{\mathrm{dyn}}$ to be $\sim2,000$\,yr, implying a recent jet ejection. We compare the jet timescale to the quasar's lifetime ($t_{\mathrm{Q}}$) that indicates the duration of the latest black hole accretion event and is derived from the proximity zone size in the rest-UV/optical spectrum. However, a ghostly Damped Ly$\alpha$ (DLA) system affects this measurement yielding an upper limit of $t_{\mathrm{Q}} \lesssim 10^4$\,yr, consistent with the jet lifetime and indicative of a young quasar. This suggests that the triggering of a UV-bright quasar phase may occur within comparable timescales as the launch of a relativistic radio jet. Therefore, we may be witnessing an early stage of black hole and jet interactions in a quasar during the first gigayear of the universe.

\end{abstract}
\section{Introduction}\label{intro}

The existence of supermassive black holes (SMBHs) with masses of $\sim 10^{8-9}$\,\Msun\ growing just within the first billion years of the universe ($z\gtrsim6$) challenges black hole formation theories \citep[e.g.][and references within]{volonteri_case_2015,volonteri_what_2023}. Several authors \citep[e.g.,][]{blandford_hydromagnetic_1982,ghisellini_role_2013,fabian_high-redshift_2014,ghisellini_sdss_2015} have proposed mechanisms through which jets could aid this growth. Recent simulations by \citet{su_self-regulation_2023,su_self-regulation_2024} use different black hole seed masses (1, 100, and $10^4$)\,\Msun\ and find that, independent of the seed mass, early jet feedback can inhibit black hole growth beyond $10^4$\,\Msun; however, sustained accretion and growth of a $10^9$\,\Msun\ SMBH by $z\sim6$ is achievable in conditions characterized by low jet velocities (3,000 km\,s$^{-1}$) and jet cocoon with high gas densities (10$^5$ cm$^{-3}$). Radio jets could thus play a key role in the formation and growth mechanisms of SMBHs within 1\,Gyr. However, calculating the lifetime of the jets, and linking them with the black hole accretion duty cycle is challenging. 

The relativistic electrons in a collimated jet suffer energy losses via synchrotron emission and cool following the power-law relationship $S_{\nu} \propto \nu^{\alpha}$, where $S_{\nu}$ is the observed flux density at the frequency $\nu$ and $\alpha$ is the radio spectral index. However, as jets traverse through the ISM, they produce shocks and exhibit magneto-hydrodynamic instabilities \citep[e.g.,][]{borse_numerical_2021}. These instabilities cause the injection of electrons in the jet to decollimate on small scales $\lesssim 5$\, kpc \citep[e.g.,][]{laing_structures_2008} into a turbulent flow of Kolmogorov characteristics \citep{kolmogorov_local_1941}. Consequently, the jet-ISM shock re-accelerates the relativistic electrons in shearing flows to high energies \citep[Lorentz factor of $\gtrsim 10^5$; ][]{bell_acceleration_1978} and then cool again following the power-law signature \citep[e.g.,][]{perucho_dissipative_2019,rieger_turbulence_2021}. Given that the electron loss timescale to synchrotron radiation is inversely proportional to its energy, higher-energy electrons radiate their energy faster than the lower-energy population. This effect is reflected as a break from a power-law pattern at low frequencies to a steeper decline at higher frequencies. 

Multiple studies on the strong radio emission of active galactic nuclei (AGN) at $z\sim0$ have modeled this synchrotron break by varying the injection of electrons and resulting spectral index $\alpha$ after the jet-ISM shock \citep[e.g.,][]{kardashev_nonstationarity_1962,pacholczyk_radio_1970,jaffe_dynamical_1973,myers_synchrotron_1985,heavens_particle_1987,meisenheimer_synchrotron_1989,carilli_multifrequency_1991,tisanic_vla-cosmos_2019,tisanic_vla-cosmos_2020}. The frequency break $\nu^{\rm break}_{\rm obs}$ above which the spectrum steepens from the injected power law relates to the \textit{spectral aging} through the parameter $t_{\mathrm{spec}}$, the cooling time experienced by the electrons since the last jet-ISM shock event. In contrast, the jet lifetime is the total time of expansion since its launch, or its \textit{dynamical age} ($t_{\mathrm{dyn}}$), which is more challenging to measure from observations given the complexity in deriving electron advance speeds as the radio jet interacts with the environment. However, numerical simulations have calculated the dynamical age from the jet and find a discrepancy to the spectral aging obtained from observations of radio galaxies at $z\lesssim 0.3$ such that $ t_{\mathrm{dyn}} \sim 1-20 \times t_{\mathrm{spec}}$ \citep[e.g.,][]{hardcastle_numerical_2013,harwood_spectral_2015,mahatma_investigating_2020}. Thus, the spectral aging sets a rough lower limit to constrain the dynamical age, or jet lifetime. Observationally, however, this phenomenon is yet to be demonstrated and quantified in sources at $z\gtrsim0.3$. 

Furthermore, we could compare the dynamical age of the jet to the black hole accretion event, as it is thought that jets arise shortly after the AGN turns on \citep[e.g.,][]{blandford_relativistic_2019}. However, it is challenging to observationally measure the lifetime of AGN. This complexity arises, in part, from the different episodes of black hole accretion. Various simulations and observational studies of quasars in the local universe show that the intermittent accretion timescales vary between a few days and $\sim 10^4$\,yr \citep[e.g.,][]{ostriker_momentum_2010,novak_feedback_2011,choi_radiative_2012}. However, the emission we observe today from the accretion disk of the quasar is being produced from only the most recent accretion episode.

It is possible to observationally measure the timescale of the latest accretion episode, the \textit{quasar lifetime, $t_{\mathrm{Q}}$}, for quasars in the epoch of reionization \citep[at $z \gtrsim5.3$,][]{bosman_hydrogen_2022}. Quasars at this cosmic epoch show almost complete absorption blueward of the \lya\ emission line due to patches of neutral hydrogen in the intergalactic medium (IGM) along the line-of-sight to the quasar. However, the quasar's UV radiation radially ionizes its immediate environment, producing an \hii\ bubble that permits enhanced flux transmission just blueward of \lya\ \citep[e.g.,][]{madau_earliest_2000,cen_quasar_2000,haiman_probing_2001,wyithe_improved_2005,keating_probing_2015,eilers_implications_2017}. This ionized bubble is bigger physically (in spatial extent) and spectrally (in terms of its effect in the IGM) for an older quasar as the surrounding medium has been ionized for a longer time. The IGM has a limited reaction time to achieve a new ionization equilibrium state as a
result of the photoionization from the source. For a typical luminous quasar with absolute magnitude at rest-frame 1,450\,$\angstrom$, $M_{1450} \sim -27$, the equilibrium time frame is $t_{eq} \approx \Gamma^{-1}_{\mathrm{H\,{\sc I}}} \approx 3 \times\ 10^4$\,yr, where $\Gamma^{-1}_{\mathrm{H\,{\sc I}}}$ is the hydrogen photoionization rate \citep[][]{eilers_implications_2017}{}{}. 

The measurement of the ionized \hii\ bubble size can be conducted following the approach in \citet{haiman_probing_2001} on \textit{Str\"omgren spheres}. However, this theory does not predict the distribution
of residual neutral hydrogen within the ionized bubble. An alternative technique developed by \citet{eilers_implications_2017,eilers_first_2018} relies on the observed size of the proximity zone ($R_{\rm p}$), defined as the extent of the transmitted flux blueward of \lya\ that drops below the 10\% level \citep[e.g.,][]{fan_constraining_2006}. This method is thus sensitive to the fraction and distribution of the residual neutral hydrogen. To calculate $R_{\rm p}$ avoiding very large uncertainties, it is essential to have a precise quasar redshift, and reliable modeling of the intrinsic continuum emission from the quasar. Radiative transfer simulations are run to understand the photo-ionization on the IGM along the line-of-sight. These compute the time-dependent ionization and recombination of different species of hydrogen and helium taking into account the associated photo-ionization heating, adiabatic cooling due to the expansion of the universe, and inverse Compton (IC) cooling of cosmic microwave background (CMB) photons \citep[e.g.,][]{bolton_nature_2007,davies_quasar_2016}. 

Different studies have observed that average sizes of proximity zones in quasars at $z\sim 5.5 - 7.5$ are $1 \lesssim R_{\rm p} \lesssim 7.1$ proper Megaparsecs (pMpc) for UV-bright quasars ($-27.5 \lesssim M_{1450} \lesssim -26.5$) corresponding to $t_{\mathrm{Q}} \gtrsim10^6$\,yr \citep{eilers_implications_2017}, while $0.5 \lesssim R_{\rm p} \lesssim 4$\,pMpc for faint quasars ($-26 \lesssim M_{1450} \lesssim -23$) implying a shorter lifetime $t_{\mathrm{Q}} <10^4$ \citep{ishimoto_subaru_2020}. However, observations of some bright quasars were unexpectedly associated with very short lifetimes of only $\sim 10^4 - 10^5$\,yr, ($0.3 \lesssim R_{\rm p} \lesssim 1.5$\,pMpc), which further challenges current black hole formation models \citep[e.g.,][]{eilers_first_2018,eilers_detecting_2021}.

The quasar PSO J352.4034--15.3373 (hereafter P352--15) is one of the most powerful radio sources known in the Epoch of reionization \citep{banados_powerful_2018,rojas-ruiz_impact_2021} hosting an extended 1.62\,kpc radio-jet \citep{momjian_resolving_2018}. \cite{rojas-ruiz_impact_2021} hereafter referred as \citetalias{rojas-ruiz_impact_2021}, used Atacama Large Millimeter/sub-millimeter Array\,(ALMA) \cii--158 $\mu$m observations to calculate the systemic redshift of the quasar $z_{sys} = 5.832\pm0.001$, crucial for the $R_{\rm p}$ measurement. Additionally, \citetalias{rojas-ruiz_impact_2021} studied the spectral energy distribution (SED) of the quasar in the rest-frame far infrared and radio. Observations from ALMA and the IRAM NOrthern Extended Millimeter Array (NOEMA) were used to model the dust continuum of the quasar host galaxy as observed at 290\,GHz and 100\,GHz, respectively. Observations from the Giant Metrewave Radio Telescope (GMRT) at 215\,MHz, and archival observations from the Very Large Array\,(VLA) at 1.4\,GHz and 3.0\,GHz were used to study the radio synchrotron emission from the jet. The results of the SED study indicated a break in the synchrotron spectrum between the observations at 3\,GHz and 100\,GHz. This implied that significant radio synchrotron emission is influencing the rest-frame far infrared continuum emission of the host galaxy likely due to the jet interacting with the ISM.

P352--15 is the first laboratory at such high redshifts enabling measurements and comparisons of the timescales of jet launching ($t_{\mathrm{dyn}}$) and black hole accretion ($t_{\mathrm{Q}}$). In this work, we investigate the frequency of the synchrotron break $\nu^{\rm break}_{\rm obs}$ to calculate the spectral and dynamical ages of the jet, $t_{\mathrm{spec}}$ and $t_{\mathrm{dyn}}$ correspondingly, based on newly presented multi-frequency observations with the VLA. We further compare the timescale of the jet launch with that of the latest black hole accretion episode, $t_{\mathrm{Q}}$, as measured from the analysis of rest-frame spectroscopic observations with the Very Large Telescope (VLT) X-shooter spectrograph \citep{vernet_x-shooter_2011}. This work is organized as follow: we describe the VLA observations and their data reduction in \S \ref{vla-obs-txt}. The VLT-X-shooter spectral observations and reduction of P352--15 are described in \S \ref{vlt-reduct}. We model the SED of P352--15, from sub-mm to radio wavelengths with data presented here and from previous work \citepalias{rojas-ruiz_impact_2021} in \S \ref{model-break}. The calculation of the dynamical age of the jet is presented in \S \ref{t_dyn-txt}. We estimate the proximity zone of P352--15 as described in \S \ref{Rp-txt} and the corresponding quasar lifetime measurement in \S \ref{ql-txt}. Finally, we compare the jet and quasar lifetime timescales and discuss caveats in our measurement and outlook for future observations in \S \ref{discussion-jet}. Throughout this work we adopt a cosmology with: H$_0$= 70 km~s$^{-1}$ Mpc$^{-1}$, $\Omega_M$= 0.3, $\Omega_\Lambda$ = 0.7, and ${T^{z=0}_{
\mathrm{CMB}}}$= 2.725 K. Using this cosmology, the age of the Universe is 948 Myr at the redshift of P352--15 and $1"$ corresponds to 5.8 proper kpc.

\section{VLA Observations and Data Reduction}\label{vla-obs-txt}
The VLA observations were performed as part of the program 21A-307 (PI: Rojas-Ruiz). Observations were taken in the L ($1-2$\,GHz), S ($2-4$\,GHz), C ($4-8$\,GHz), X ($8-12$\,GHz), Ku ($12-18$\,GHz), K ($18-26.5$\,GHz), Ka ($26.5-40$\,GHz), and Q ($40-50$\,GHz) bands to construct the radio SED of P352--15 and to determine the exact location of a break or a turnover in the synchrotron emission. 

\begin{table}
\begin{center}
\caption{Summary of VLA Observations of P352--15 \label{vla-obs}}
\begin{tabular}{lccccc}
\hline
\hline
Band & Flux Cal. & Gain Cal. & On-target & Array\\
 & &  & time (s) & Config.\\
\hline
L & 3C147 & J2348--1631 & 336 & D\\
S & 3C147 & J2348--1631 & 326 & D\\
C & 3C147 & J2331--1556 & 325 & D\\
X & 3C147 & J2331--1556 & 325 & D\\
Ku & 3C147 & J2331--1556 & 357 & C\\
K & 3C147 & J2331--1556 & 390 & C\\
Ka & 3C147 & J2331--1556 & 387 & C\\
Q1 & 3C147 & J2331--1556 & 2418 & C\\
Q2 & 3C147 & J2331--1556 & 1860 & C\\

\hline
\end{tabular}
\end{center}
\end{table}

We designed our observations by assuming a break near 3\,GHz (the most conservative case) and aiming to detect the expected emission with a signal-to-noise ratio S/N$>15$. This S/N also allows us to determine an intra-slope for each receiver band, which is important to find an accurate location of the turnover frequency. Since we know that the emission of our source comes from a region smaller than 0\farcs5 \citep{momjian_resolving_2018} and there are no other strong radio sources in the field \citep{banados_powerful_2018}, we used the compact C and D configurations. We note that data of P352--15 at NVSS 1.4\,GHz and VLA DDT 3\,GHz were already available in the literature \citep{banados_powerful_2018}, but we repeated observations at such frequencies to discard any intrinsic quasar variability effect. 

The observations are quasi-simultaneous spanning $\sim$5 months ---between 2021 March 26 and 2021 August 17--- which corresponds to less than a month in the quasar rest-frame (given the $1+z$ time delay). The total allocated time was 5 hours, of which 1 hour and 52 minutes were on source. Table \ref{vla-obs} lists the flux density scale calibrator used in these observations as well as the complex gain calibrator for each frequency band, the total integration time on the target source, and the array configuration.

The observations were taken in four parts. The first set of observations was performed in the L, S, C, and X bands on 2021 March 26 in the D configuration. L and S-band observations were taken with the 8-bit samplers offering a bandwidth of 1\,GHz for L-band and 2\,GHz for S-band.
The C and X-band observations were taken with the 3-bit samplers totaling 4\,GHz of bandwidth per band. The next set of observations was taken on 2021 June 11 in the C configuration at Ku and K bands and utilized the 3-bit samplers. To achieve the desired sensitivity at the Q-band, a separate single-band observing session was carried out. The first set referred to as Q1 in Table \ref{vla-obs} was taken on 2021 July 31 in C-configuration using the 3-bit samplers. Combined observations in the Ka and Q (Q2 in Table \ref{vla-obs}) bands were carried out on 2021 August 17 in C configuration also using the 3-bit samplers.

\begin{table}
\begin{center}
\caption{Flux Measurements of P352--15 in VLA Bands\label{vla-tab-data}}
\begin{tabular}{ccccc}
\hline
\hline
Band & $\nu_{\mathrm{cent.}}$ & $S_\nu$ & S/N & Beam \\
{} & (GHz)  & (mJy) & {} & arcsec\\
\hline
L & 1.5 & 14.89$\pm$3.73& 3.9 & 51.3$\times$27.6\\
S1 & 2.5 & 9.41$\pm$0.09& 20 & 38.5$\times$20.0\\
S2 & 3.5 & 6.93$\pm$0.05& 20  & 34.1$\times$15.5\\
C1 & 4.5 & 5.52$\pm$0.03& 20 & 25.4$\times$13.3\\
C2 & 5.5 & 4.67$\pm$0.03& 20 & 21.8$\times$11.3\\
C3 & 6.5 & 3.78$\pm$0.03& 20 & 18.4$\times$9.5\\
C4 & 7.5 & 3.32$\pm$0.03& 20 & 16.5$\times$8.4\\
X1 & 8.5 & 2.80$\pm$0.02& 20 & 14.3$\times$7.1\\
X2 & 9.5 & 2.51$\pm$0.02& 20 & 13.2$\times$6.5\\
X3 & 10.5 & 2.25$\pm$0.03& 19 & 12.3$\times$6.1\\
Ku1 & 12.8 & 1.78$\pm$0.04& 19  & 3.2$\times$1.5\\
Ku2 & 14.3 & 1.67$\pm$0.03& 19 & 3.1$\times$1.3\\
Ku3 & 15.8 & 1.52$\pm$0.03& 19  & 2.7$\times$1.2\\
Ku4 & 17.4 & 1.37$\pm$0.03&  18 & 2.1$\times$1.1\\
K1 & 19.1 & 1.19$\pm$0.03& 9.7 & 2.5$\times$1.0\\
K2 & 21.0 & 1.05$\pm$0.03& 9.6 & 1.8$\times$0.9\\
K3 & 23.0 & 1.02$\pm$0.04& 9.4 & 1.6$\times$0.8\\
K4 & 25.0 & 0.84$\pm$0.03& 9.4 & 1.5$\times$0.8\\
Ka1 & 30.1 & 0.76$\pm$0.06& 6.8 & 1.3$\times$0.6\\
Ka2 & 32.0 & 0.72$\pm$0.06& 6.6 & 1.2$\times$0.6\\
Ka3 & 34.0 & 0.61$\pm$0.06& 6.2 & 1.1$\times$0.5\\
Ka4 & 36.0 & 0.65$\pm$0.08& 5.7 & 1.1$\times$0.5\\
Q & 42.0 & 0.46$\pm$0.04& 5.8 & 1.0$\times$0.5\\

\hline
\end{tabular}
\end{center}
\end{table}

\subsection{Data Reduction Process}\label{vla-reduce}
All the data were processed using the Common Astronomy Software Applications package \citep[CASA,][]{mcmullin_casa_2007} for calibrations after manually excising bad data, as well as deconvolution and imaging. We edited out channels heavily impacted by radio frequency interference (RFI).\footnote{\url{https://science.nrao.edu/facilities/vla/docs/manuals/obsguide/rfi\#section-3}} For all bands, the data were self-calibrated in order to correct visibility phase and amplitude errors from sources in the field, particularly in the lowest-frequency bands L, S, C, X.

L-band data were cleaned using the Hogbom algorithm and Briggs weighting \citep{briggs_high_1995} with a robust parameter of 0.4. The S-band data were split into two frequency chunks, each one 1\,GHz wide (see central frequencies in Table \ref{vla-tab-data}). The data were cleaned with the deconvolver `mtmfs' and 2 terms of Taylor coefficients for the spectral model. Imaging was performed with Briggs weighting and a robust parameter value of 0.4. The C-band was divided into 4 frequency chunks of 1\,GHz wide, each imaging was produced using the `mtmfs' algorithm and Briggs weighting with a 0.8 robust value. The process of calibrating and imaging was the same for the X-band but in this case, the last frequency chunk was dominated by RFI and thus was discarded. See images in Appendix \ref{Appendix-vla1}.

The Ku, K, and Ka bands had minimal RFI contamination. Each band was split into four frequency chunks per observation with band widths of 1.5\,GHz, 2.0\,GHz, and 2.0\,GHz, correspondingly. All chunks were imaged with the `mtmfs' algorithm and Briggs weighting with a robust value of 0.8. One Q session (Q2) showed anomalies compared to the other (Q1) session and the expected spectral behavior compared to the Ka-band. Thus, these were discarded.
The final imaging from the Q-band is from the Q1 observations, which have longer exposure time as well. Given the lower S/N in this band, the source is only detected in the frequency range of (40 -- 44\, GHz), and thus we do not divided the data into shorter frequency chunks. The final images from Ku, K, Ka, and Q observations are presented in Appendix \ref{Appendix-vla1}.

For all images, the flux density was measured by fitting a 2D Gaussian with CASA, and the root-mean-square (rms) noise was measured from the image rather than the fit of the Gaussian in order to account for all noise variations around the source. Table \ref{vla-tab-data} shows the central frequency for each VLA frequency chunks described above, the flux density, the signal-to-noise (S/N) ratio calculated using the rms noise, and the synthesized beam size. These measurements are used to model the synchrotron spectrum of P352--15, as shown in Figure \ref{sync-models}. 

\section{VLT-X-shooter Spectral Observations and Data Reduction}\label{vlt-reduct}
Here we describe the observations necessary to measure the proximity zone $R_{\rm p}$ for the quasar P352--15, to later calculate the \textit{quasar lifetime} $t_{\mathrm{Q}}$.

Rest-UV/Optical spectral data of P352--15 were taken with VLT/X-shooter \citep{vernet_x-shooter_2011} as part of program ID = 0102.A-0931 (PI: B. Venemans). X-shooter is a multi-wavelength, medium-resolution spectrograph mounted at the VLT UT2 telescope. It consists of three spectroscopic arms observing in the UVB spectrum (300--559.5\,nm), VIS (559.5--1024\,nm), and NIR (1024--2480\,nm). Given the Gunn-Peterson trough effect, we only obtain spectral information from the quasar P352--15 at $z=5.832$ in the VIS and NIR arms. These data were taken on 2018 October 3 and 11. For the NIR observations, nodding along the slit in an ABAB pattern was used for taking offset exposures, which is later used as a double pass sky subtraction.

The data reduction is performed with \texttt{PypeIt}\footnote{\url{https://pypeit.readthedocs.io/en/release}}\ developer version 1.9.2 \citep{prochaska_pypeit_2020}. The VIS and NIR data are similarly reduced with standard techniques. The two sets of science frames from both dates of observations are used conjunctively during the reduction process. The first calibrations involve sky subtraction on the 2D images (including the difference imaging of dithered AB pairs of exposures for the NIR arm case) and applying a 2D b-spline fitting procedure. Object traces from the 2D images are identified and extracted to produce the 1D spectra following the optimal spectrum extraction technique from \citet{horne_optimal_1986}. Next, the 1D spectra for VIS and NIR are flux calibrated using the GD71 star. The resulting fluxed 1D spectra are coadded and corrected for telluric absorption. This last correction is performed using the  \texttt{PypeIt}-provided telluric file from Paranal Observatory\footnote{\url{https://pypeit.readthedocs.io/en/release/telluric.html}}, which identifies the wavelength ranges of the tellurics, and applies telluric model grids produced from the Line-By-Line Radiative Transfer Model \citep[LBLRTM\footnote{\url{http://rtweb.aer.com/lblrtm.html}};][]{clough_atmospheric_2005,gullikson_correcting_2014}. The telluric features are then corrected to later join the VIS and NIR final spectrum of P352--15. In \S \ref{Rp-txt}, we report and use the spectrum normalized at the 1290\AA\ flux density.

\section{Calculating the Frequency Break of the Synchrotron Emission}\label{model-break}
In order to correctly model the synchrotron spectral break $\nu^{\rm break}_{\rm obs}$ of P352--15, we must combine the models for the cold dust with available observations from \citetalias{rojas-ruiz_impact_2021}, and synchrotron emission with the new VLA data.

\subsection{Modeling the Cold Dust from mm Observations}
We use the mm observations from \citetalias{rojas-ruiz_impact_2021} at 290\,GHz and 100\,GHz to model the cold dust with a modified black body (MBB)  \citep[see, e.g.,][]{venemans_dust_2018,novak_alma_2019} of the form:

\begin{equation}
    S_{\rm obs} = f_{\rm CMB} \ \left[1 + z\right] \ D_L^{-2} \  \kappa_{\nu_{\mathrm{rest}}}(\beta) \ 
     M_{\mathrm{Dust}} \ B_{\nu_{\mathrm{rest}}}(T_{\mathrm{Dust,z}})
     \label{mod-bb}
\end{equation}

Here, $f_{\rm CMB}$ is a correction against the CMB contrast (see \citealt{da_cunha_effect_2013}), $D_L$ is the luminosity distance, $\kappa_{\nu_{\mathrm{rest}}}(\beta)$ is the dust mass opacity which depends on the dust emissivity spectral index $\beta$ (see \citealt{dunne_scuba_2000,dunne_census_2003,james_scuba_2002}), \Mdust is the dust mass and $B_{\nu_{\mathrm{rest}}}(T_{\mathrm{Dust,z}})$ is the Planck function in terms of frequency that depends on the host galaxy dust temperature and redshift. For appropriate calculation, all values should be converted to the same system of units.

\citetalias{rojas-ruiz_impact_2021} calculated the dust mass of P352--15's host galaxy using the continuum underlying the \cii-158$\mu$m emission --observed with ALMA at 290\,GHz and shown to not be significantly contaminated by synchrotron emission-- to be \Mdust $=(0.36\pm 0.04)\times 10^8$ $M_\odot$, and presented a range of acceptable values for the dust temperature T $=30-100$\,K and emissivity index $\beta=1.6 - 1.95$. Typical cold dust parameters of high-redshift quasars are T$_{\rm{dust}}$ = 47 K and $\beta= 1.6$ \citep[e.g.][]{beelen_350_2006,venemans_bright_2016}, however, the lack of information in the mm and radio SED prevented accurate calculations of these parameters due to overfitting. On this occasion, however, given the VLA supplementing data, we leave T$_{\rm{dust}}$ and $\beta$ as free parameters to fit a joint MBB and synchrotron break model.

\begin{figure*}[hbpt!]
\centering
\includegraphics[width=0.85\linewidth]{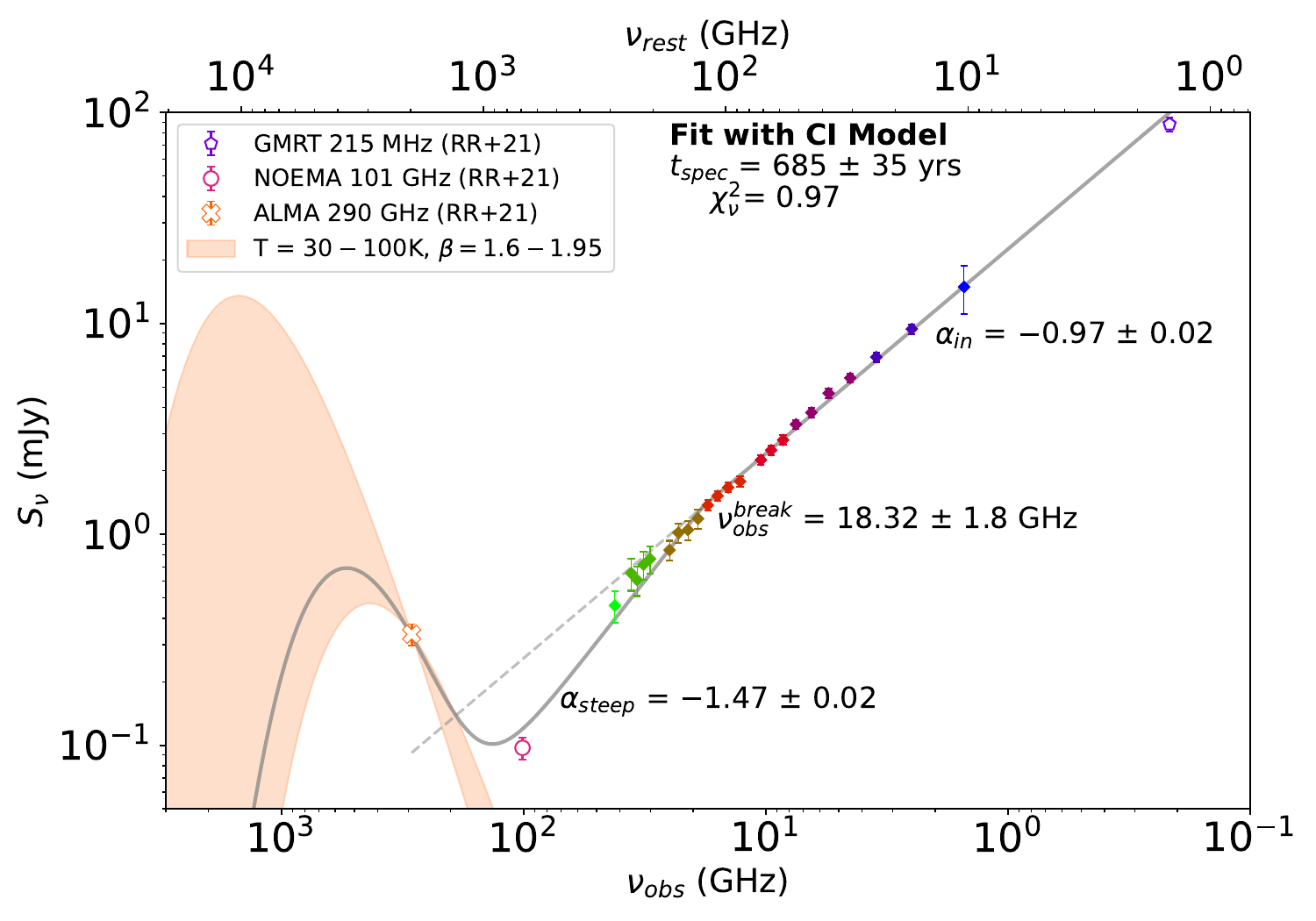}
\includegraphics[width=0.85\linewidth]{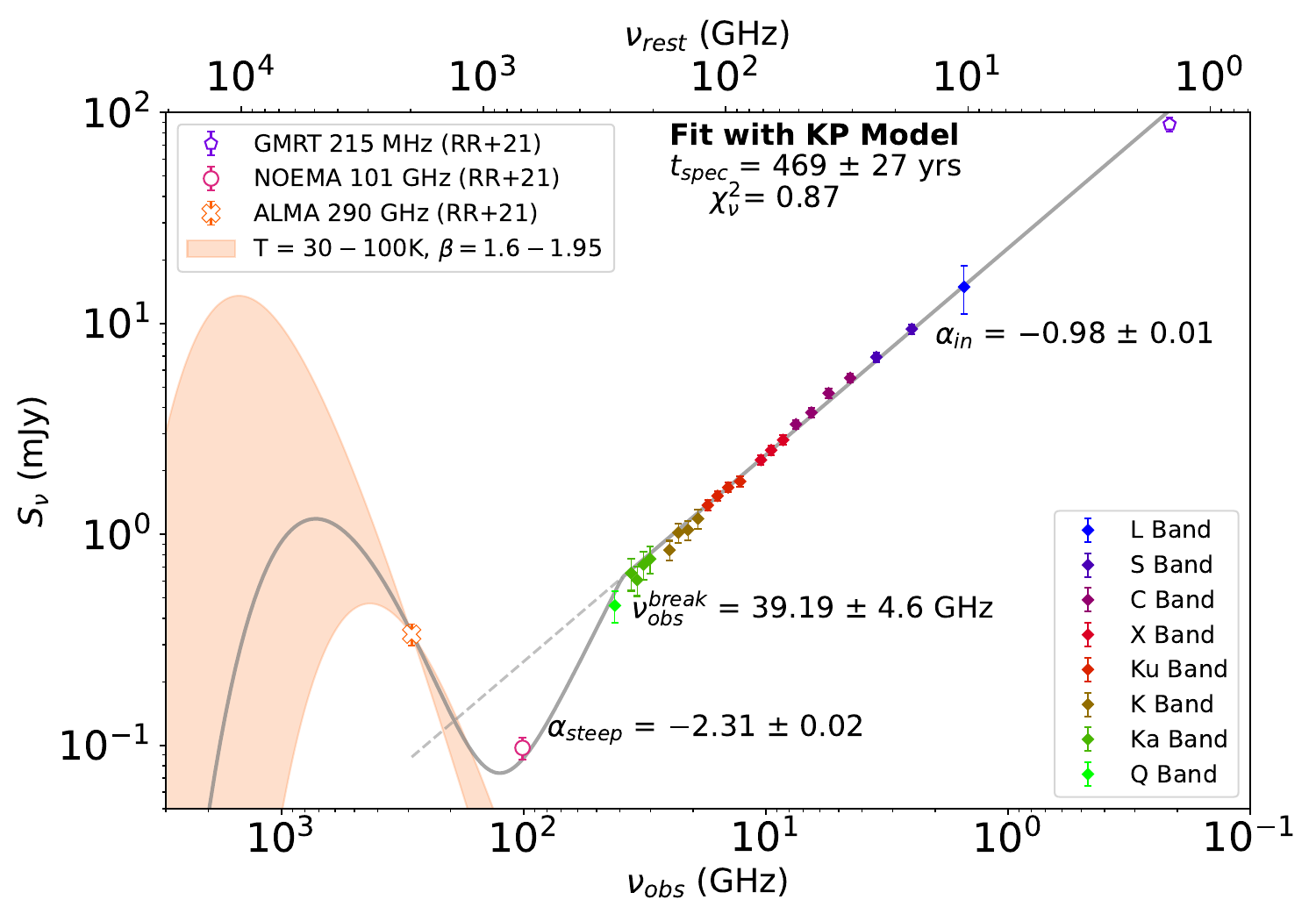}

\caption[The mm to radio SED fit of P352--15 based on different synchrotron break models.]{The mm to radio SED of P352--15. The mm-observations presented here are from \citetalias{rojas-ruiz_impact_2021}, and the radio observations are from this work in all the VLA from 1.5\,GHz - 42\,GHz. We report the combined fit of the cold dust emission, modeled with an MBB, and a synchrotron break, modeled with the CI \textit{(top)} and KP \textit{(bottom)} models (see \S \ref{model-break}). We note that both fits are consistent with our observations, the difference relies on the assumption of relativistic electron distribution. The resulting spectral aging of the jet $t_{\mathrm{spec}}$ from both models falls within the same order of magnitude and for the rest of this work we consider both values as upper and lower estimates, correspondingly.}
\label{sync-models}
\end{figure*}

\subsection{Modeling the Synchrotron Spectral Break}
The intrinsic synchrotron emission
is well-described by the power law relationship between the flux density and spectral index $S_{\nu} \propto \nu^{\alpha}$.
A break in the power law at high frequencies
has been well characterized in the past using observations and theoretical models on jet physics, particle distribution, and energy loss. The models we use to fit the break of the synchrotron spectrum are described as follows: 
\begin{itemize}
\item The continuous injection model (abbreviated as CI) was initially described
in \citet{heavens_particle_1987} and then applied to observations of hot spots (where the jet interacts with the ISM) in \citet{meisenheimer_synchrotron_1989}. This model involves a constant injection rate of relativistic electrons in the power law distribution. This case simulates an observed unresolved source where different populations of electrons are seen at once \citep{carilli_multifrequency_1991}, as in our VLA observations of P352--15. In this model, the power law index above the frequency break $\nu^{\rm break}_{\rm obs}$ ($\alpha_{\mathrm{steep}}$) relates to the injected power law index ($\alpha_{\mathrm{in}}$) as: $\alpha_{\mathrm{steep}} =  \alpha_{\mathrm{in}} - 0.5$.

\item The Kardashev-Pacholczyk (KP) model described in \citet{kardashev_nonstationarity_1962,pacholczyk_radio_1970} assumes a ``one-shot'' injection of relativistic electrons into the power law distribution. The particles will maintain their initial pitch angle with respect to the magnetic field over time i.e., no scattering. After the frequency break $\nu^{\rm break}_{\rm obs}$, the new power law depends on the injected spectral index in the following relationship $\alpha_{\mathrm{steep}} = (\frac{4}{3})\alpha_{\mathrm{in}} - 1$. This model is often preferred to describe sources with strong hot spots \citep[e.g., FRII sources,][]{harwood_spectral_2013,harwood_spectral_2015}. The VLBA observations of P352--15 in \citet{momjian_resolving_2018} reveal strong hot spots.
\end{itemize}

The Jaffe-Perola (JP) model \citep{jaffe_dynamical_1973} was also considered for modeling our data but was ultimately dismissed. This model also incorporates a ``one-shot'' injection of electrons but allows for scattering of the pitch angle with time. This timescale is still shorter than the radiative lifetime of the particles and thus results in a sharp exponential cutoff to the power law distribution of particles. Given that there is a gap in the frequency coverage between the VLA Q-band at 42\,GHz and the 100\,GHz observation from NOEMA in \citetalias{rojas-ruiz_impact_2021}, this exponential drop cannot be observed and thus we do not include the JP model in the analysis of this work.

Finally, we could in principle make a ``free model'' as in the case of \citet{tisanic_vla-cosmos_2019,tisanic_vla-cosmos_2020} to fit a totally free double power-law, i.e. without constraints on the steepened power law $\alpha_{\mathrm{steep}}$. Applying this free model to our data yielded sub-optimal parameter estimation. However, the resulting spectral aging $t_{\mathrm{spec}}$ measurement was consistent within 1$\sigma$ of the mean spectral aging obtained using the well-studied CI and KP models, both of which account for jet physics and are suitable for describing the synchrotron emission of P352--15 (see more on the spectral aging estimates in \S \ref{t_dyn-txt}).

In Figure \ref{sync-models}, we show the mm and radio observations of P352--15 fitted with a joint function of an MBB and synchrotron steepening model, assuming the CI (\textit{top}) and the KP (\textit{bottom}) modeling. Following \citet{shao_radio_2022}, we additionally considered a calibration error intended for performing this joint fit using the VLA observations. Adhering to the recommendations of the National Radio Astronomy Observatory (NRAO), we account for flux density inaccuracies of 5\% for bands L through Ku, 10\% for the K-band, 12.5\% for the Ka-band, and 15\% for the Q-band. The free parameters to fit with this function are the dust temperature ($T$) and emissivity index ($\beta$), the spectral index of the injected power law ($\alpha_{\mathrm{in}}$), a constant ($\omega$) relating the power law proportionality $S_{\nu} =\omega \times \nu^{\alpha}$, and the frequency break ($\nu^{\rm break}_{\rm obs}$). 

Assuming the CI model, the reduced chi-squared of the fit to the MBB + synchrotron break function is $\chi^2_\nu=0.97$. The frequency break of the synchrotron emission is found to be $\nu^{\rm break}_{\rm obs}=18.32\pm1.8$\,GHz, with an injection power law slope $\alpha_{\mathrm{in}} = -0.97\pm0.02$ and a corresponding steepened slope after the break of $\alpha_{\mathrm{steep}}= -1.47\pm0.02$. Adopting the KP model results in a similarly good fit with a comparable reduced chi-squared of $\chi^2_\nu=0.87$. In this case, the frequency break occurs at nearly double the value obtained from the previous model such that $\nu^{\rm break}_{\rm obs}=39.19\pm4.6$\,GHz. This can be explained because the KP model describing a ``one-shot'' injection of relativistic electrons results in a steeper slope after the synchrotron break, where in this case $\alpha_{\mathrm{in}} = -0.98\pm0.01$ and $\alpha_{\mathrm{steep}} = -2.31\pm0.02$. 

We note that the dust parameters $T$ and $\beta$ corresponding to the MBB are still not well-constrained given the lack of information on the cold dust emissivity at higher frequencies of the 290\,GHz ALMA observation. Additional sub-millimeter observations would be required to refine the modeling of the MBB peak and provide a more accurate characterization of the cold dust emissions from the quasar host galaxy. However, we can robustly conclude from the SED fits that the observed 290\,GHz continuum near \cii\ is not contaminated by synchrotron emission.

\section{Calculating the Jet Timescale}\label{t_dyn-txt}

The evolution of relativistic particles in jets and hot spots, involving interactions with the ISM, has been extensively studied, with \citet{leahy_interpretation_1991} linking the radiative aging of electrons and the spectral aging $t_{\mathrm{spec}}$ using the equation:

\begin{equation}\label{tspec-eqn}
    t_{\mathrm{spec}}=50.3 \frac{B^{1 / 2}}{B^2+B_{\mathrm{CMB}}^2}\left((1+z) \nu^{\rm break}_{\mathrm{rest}}\right)^{-1 / 2} \ [\mathrm{Myr}]
\end{equation}
\noindent where $B$ is the magnetic field of the jet, $B_{\mathrm{CMB}}$ is the equivalent magnetic field of the CMB at the redshift $z$ of the object, and $\nu^{\rm break}_{\mathrm{rest}}$ is the rest-frame frequency break. In this equation, magnetic fields are in units of nano Tesla (nT), the frequency is in GHz, and the aging time has units of Myr. The magnetic field strength for the jet of P352--15 was previously estimated as $B\sim3.5$\,mG (or $B\sim350$\,nT) from observations with the Very Long Baseline Array \citep[VLBA,][]{momjian_resolving_2018}. Adopting this $B$ and the frequency breaks found in the previous section, we find that $t_{\mathrm{spec}} = 685 \pm 35$\,yr, and $t_{\mathrm{spec}} = 469 \pm 27$\,yr assuming the CI and KP models of the power law break, respectively. Note that the quoted uncertainties are only statistical, and do not reflect the larger systematic uncertainty in the magnetic field strength, which is calculated using minimum energy arguments (see more details in \citealt{momjian_resolving_2018}).

Measuring this radiative aging of electrons does not necessarily represent the timescale of the jet since its launch, which is defined as the dynamical age $t_{\mathrm{dyn}}$. Discrepancies between $t_{\mathrm{spec}}$ and $t_{\mathrm{dyn}}$ may arise from the changing strength of the magnetic field as the electrons dissipate along the jet and other environmental effects involving the ISM and jet hot spots \citep[see more e.g, in][]{rudnick_relativistic_1994,jones_simulating_1999}. In fact, 
$t_{\mathrm{spec}}$ is found to be consistently shorter than $t_{\mathrm{dyn}}$ given that the methodology used to calculate the spectral aging relies more heavily on the magnetic field variation rather than the intrinsic synchrotron aging \citep[e.g.,][]{kaiser_environments_2000,machalski_multifrequency_2009}.

Several approaches have been introduced to compensate for the uncertainty in calculating dynamical ages from spectral ages that account for effects including kinetic energy input into the jet from the AGN, synchrotron emission in different areas of the jet, and shocks with the ISM \citep[e.g.,][and reference within]{machalski_multifrequency_2009}. However, \citet{blundell_spectra_2000} studied the differences among the two timescales and concluded that, if $t_{\mathrm{spec}}\ll10^7$\,yr--as it is in this case--then one can use the spectral aging $t_{\mathrm{spec}}$ and observed synchrotron frequency break $\nu^{\rm break}_{\rm obs}$ to approximate the measurement of the dynamical age $t_{\mathrm{dyn}}$. 
We derive $t_{\mathrm{dyn}}$ by first evaluating the cooling time of electrons with Lorentz factor $\gamma_e$ in the fluid frame. The cooling occurs through synchrotron loss measured by the power radiated by electrons when interacting with the jet magnetic field. We also account for inverse Compton losses with CMB interactions. The total cooling rate in the fluid frame is transformed to the observer's frame through the bulk Lorentz factor $\Gamma_{\rm jet}$ and the observed synchrotron frequency break $\nu^{\rm break}_{\rm obs}$. This yields the expression:

\begin{multline}\label{eq-tdy}
t_{\mathrm{dyn}}\left(\nu_{\mathrm{obs}}\right)=\frac{3}{4 \sigma_T\left(U_B+U_{\mathrm{CMB}}\right)}\\
  \times
\left(\frac{m_e c e B}{2 \pi \Gamma_{\rm jet}}\right)^{1 / 2} 
\left(\frac{1+z}{\nu^{\rm break}_{\rm obs}}\right)^{1 / 2}\ [\mathrm{kyr}]
\end{multline}

\noindent In this equation, $\sigma_T$ is the Thompson cross section, $U_B$ is the jet's magnetic energy density, $U_{\rm CMB}$ is the CMB energy density at the redshift $z$ of the source, $m_e$ is the electron mass, and $c$ is the speed of light. These five parameters are given in centimeter--gram--second (cgs) units. The electron charge $e$ is in electrostatic unit of charge (esu units). The magnetic field of the jet $B$ is in units of Gauss, $\Gamma_{\rm jet}$ is the bulk Lorentz factor of the jet, and $\nu^{\rm break}_{\rm obs}$ is the observed synchrotron frequency break in Hz. 

The bulk Lorentz factor of the jet ($\Gamma_{\rm jet}$) 
is calculated following:

\begin{equation}\label{Gammajet}
    \Gamma_{\rm jet}=[1-\beta^2_{\rm jet}]^{-1/2}  
\end{equation}

where $\beta_{\rm jet} = v_{\rm jet}/c$ with $v_{\rm jet}$ estimated using the jet's observed spatial extension ($\approx1.62$\,kpc, \citealt{momjian_resolving_2018}) and the estimated electron travel time that can be approximated as $10\times \langle t_{\mathrm{spec}} \rangle$, following previous studies \citep[e.g.,][]{mahatma_investigating_2020}. We estimate the spectral age of the jet by calculating the average of the synchrotron frequency break previously found with the CI and KP models, $\langle \nu^{\rm break}_{\rm obs} \rangle \sim 28.76$\,GHz, yielding $\langle t_{\mathrm{spec}} \rangle \sim 580$\,yr. The resulting bulk velocity is $\beta_{ \rm jet} \approx0.90$, corresponding to a bulk Lorentz factor $\Gamma_{\rm jet}\approx2.35$. 

\begin{figure*}[t!]
\centering
\includegraphics[width=\linewidth]{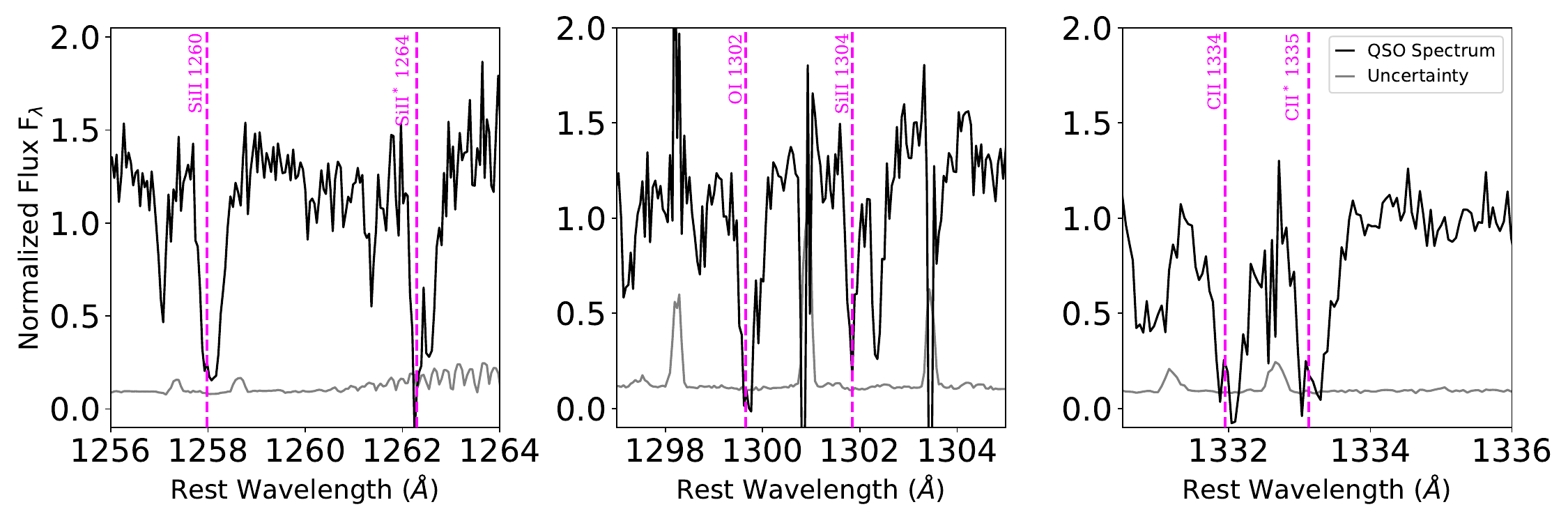}

\caption{The rest-frame spectrum of P352--15 normalized at rest-frame $1290\pm2.5$\AA\ is shown in black, and its uncertainty in grey. Each panel is a zoom-in of the quasar spectrum to show examples of the strong low-ionization absorption lines (magenta) from the system consistent with a redshift of $z_{abs}=5.8188$. The left panel shows the absorption features by \siii\, $\lambda\,1260$ and \siii$^* \lambda$\,1264, the middle panel shows the \oi\,$\lambda$ 1302, and \siii\,$\lambda$ 1304 absorption, and the right panel the \ciiS\,$\lambda$ 1334 and \ciiS$^*\lambda$ 1335 absorption. 
}
\label{low_ion}
\end{figure*}

We finally calculate the dynamical age of the jet applying Equation \ref{eq-tdy} and using $\Gamma_{\rm jet}=2.35$, the magnetic field of the jet $B\sim3.5$\,mG, and the average frequency break from the CI and KP model fits $\langle \nu^{\rm break}_{\rm obs} \rangle \sim 28.76$\,GHz. The resulting dynamical age of P352--15 is of only $t_{\mathrm{dyn}}\sim2,000$\,yr. This is the best approximation we can report given the large uncertainty in the magnetic field ($B$), which is three orders of magnitude stronger than typically observed magnetic fields in nearby radio-loud\footnote{Radio-loudness is defined as a source with a ratio $R>10$ of flux densities in rest-frame radio at 5\,GHz to the UV at 2500\,$\angstrom$, $R_{2500}=S_{5\mathrm{GHz}}/S_{2500\,\angstrom}$  \citep{sramek_radio_1980}, or to the rest-frame optical 4400\,$\AA$, $R_{4400}=S_{5\mathrm{GHz}}/S_{4400\,\angstrom}$  \citep{kellermann_vla_1989}} galaxies \citep[e.g., $B \sim 45-65\mu$G;][]{carilli_multifrequency_1991}. Finally, the uncertainties in our calculated synchrotron spectral break $\langle \nu^{\rm break}_{\rm obs} \rangle$ are statistical rather than systematic. We further discuss the uncertainties in the jet age determination in \ref{uncert-jet-age}.

Since jets are thought to be launched soon after the black hole begins accreting \citep[e.g.,][]{blandford_relativistic_2019}, we compare this jet timescale to the actively accreting black hole timescale by constraining the past history of the ionizing UV emission via the proximity zone, or to the quasar lifetime $t_{\mathrm{Q}}$.

\section{Measuring the Proximity Zone}\label{Rp-txt}
We analyzed the X-shooter spectrum of P352--15 with the \texttt{linetools}\footnote{\url{https://linetools.readthedocs.io/en/latest/xspecgui.html}} Python package in order to check the quality of the reduction process. We checked for any absorbers in the line-of-sight presented in the spectra, primarily searching for possible Lyman-limit systems (LLS) or Damped Lyman-$\alpha$ (DLA) systems that could prematurely truncate the proximity zone and compromise the lifetime measurement \citep[e.g.,][]{eilers_implications_2017,eilers_detecting_2020,banados_metal-poor_2019}.
Indeed, we found absorption systems at various redshifts in the quasar spectrum, in particular, we identified a $z_{abs}=5.8188$ absorber, which had been previously observed in \citet{banados_powerful_2018}. We find that this system has several strong high and low ionization metal lines (\nv$\,\lambda \lambda$1238,1242; \siii$\,\lambda \lambda$1260,1264; \oi$\, \lambda$1302; \siii$\,\lambda$1304; \ciiS $\, \lambda \lambda$1334,1335; \siiv\,$\lambda \lambda$1393,1402; \civ\,$\lambda \lambda$1548,1550; \alii\,$\lambda$1670; see Figure~\ref{low_ion}). This system falls in the middle of the quasar proximity zone.

Low-ionization absorbers are usually associated with large \hi\ column densities of $N_{HI}\gtrsim 10^{20}$\,cm$^{-2}$ \citep[e.g.,][]{poudel_early_2018,cooper_heavy_2019,simcoe_interstellar_2020}. Thus, the absorber should block the ionizing emission from the quasar, preventing the proximity zone from existing at lower redshifts than the absorber's ($z < z_{abs}$). The high column density should give rise to a broad fully-absorbed trough around $z_{Ly\alpha} \sim z_{abs}$. 
This characteristic column density from the absorber would be too broad to allow the presence of flux transmission from the quasar ionizing the IGM as visible in the blueward region of the \lya\ of P352--15, thus truncating the proximity zone.
However, we observe substantial flux at the line position and further blueward of the spectrum, as detailed in \S \ref{pca-section}. 
This partial transmission effect suggests that only a portion of the quasar's emission is being blocked by the absorber. Thus, this system must be physically associated with the quasar sub-pc environment, i.e., it could be a small ``metal cloud'' outflowing from the quasar and present on the scales of the broad-line region \citep[BLR; $\sim0.01 - 0.1$\,pc; e.g,][]{suganuma_reverberation_2006,cackett_reverberation_2021}, rather than a typical foreground DLA system at a lower redshift. The observation of strong \ciiS$^*\lambda$\,1335 absorption supports this interpretation, as this line is extremely rare and typically arises in a very high ionizing field \citep{bosman_fluorescent_2019}, such as that of the BLR.

\begin{figure*}[t!]
\centering
\includegraphics[width=\linewidth]{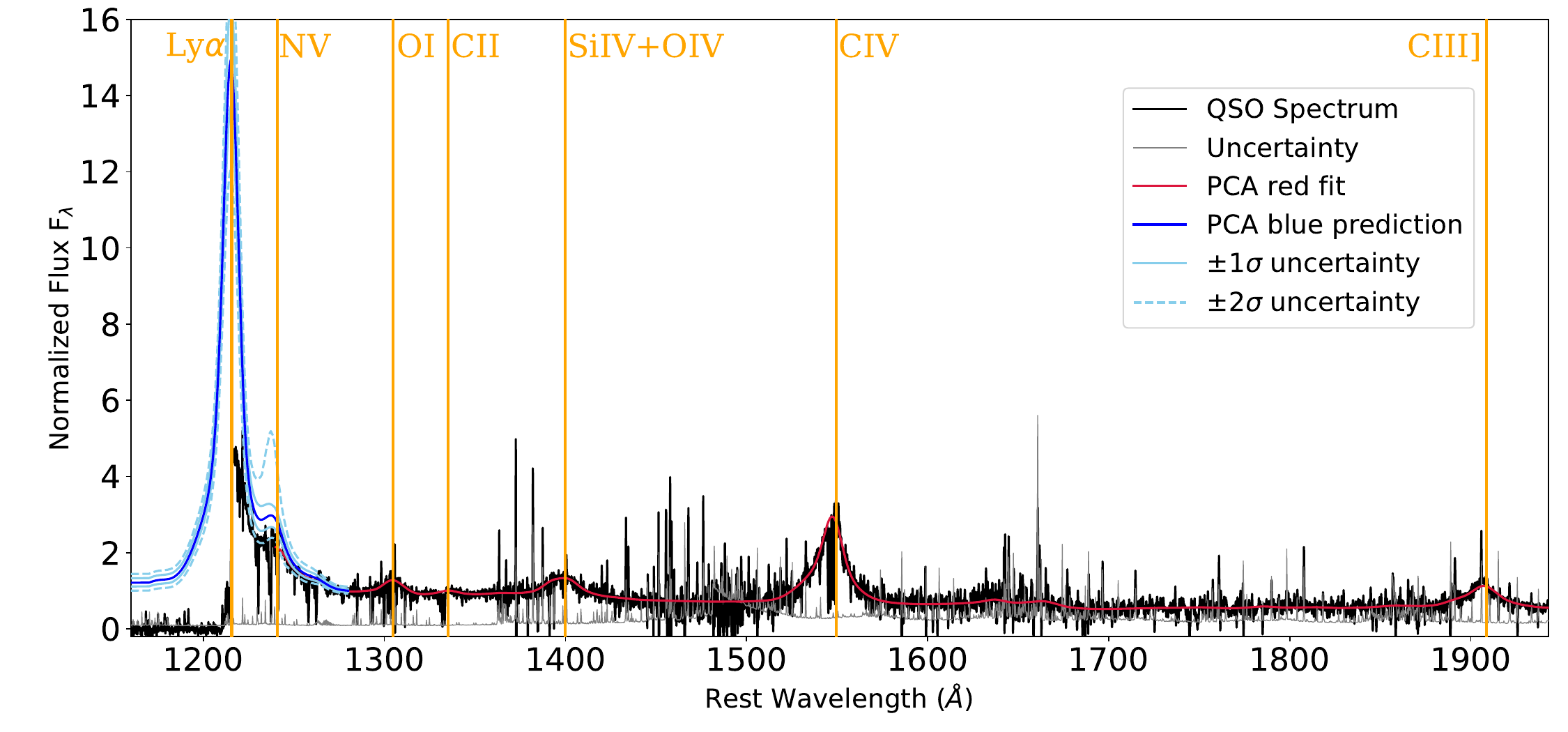}

\caption[The rest-UV/Optical spectrum of P352--15 with the PCA fit and prediction for the continuum.]{The spectrum of P352--15 normalized at rest-frame $1290\pm2.5$\AA\ is shown. A boxcar smoothing of four pixels is applied to both the measured flux (black) and uncertainty (grey). The anticipated emission lines at rest-frame wavelengths are represented by yellow vertical lines. The red-side PCA continuum fit (red) is performed by first removing any absorption features with a cubic-spline procedure. The PCA prediction of the blue-side continuum of the spectrum  ($\lambda<1290$\,\AA) is the blue line with $\pm1\sigma, \pm2\sigma$. The predicted continuum suggests a strong \lya\ damping wing with only a third of the emission present close to line center.
}
\label{spec-pca}
\end{figure*}

These types of absorbers in the sub-pc environment of quasars have been recently discovered in just a handful of low-redshift quasars, which have been christened `ghost-DLA system' \citep[e.g.,][]{fathivavsari_ghostly_2017,fathivavsari_ghostly_2020}. One example intriguingly happens to be
a radio-loud quasar at $z=2.1$ with evidence of jet-ISM interaction \citep{gupta_h_2022}. In summary, the partial absorption seen in the spectrum of P352--15 still allows flux transmission from the quasar to be visible. Also, there is no truncation at $z < z_{abs}$ implying that we are still seeing foreground IGM over-ionized by the quasar. Consequently, we continue to measure an upper limit on the proximity zone $R_{\rm p}$ of P352--15.

\subsection{Continuum Emission Prediction using PCA}\label{pca-section}

The proximity zone $R_{\rm p}$ is defined as the region of the spectrum with transmission above 10\% of the flux normalized by the continuum of the quasar. Therefore, it is necessary to predict the intrinsic quasar continuum profile of the \lya\ emission. This continuum prediction is possible
given the naturally strong correlation of the broad emission lines of a quasar's spectrum in the rest-frame UV \citep[e.g.,][]{francis_objective_1992,yip_spectral_2004,suzuki_quasar_2006,greig_lya_2017}.
Therefore, it is possible to model the emission lines redward of \lya\ to predict its intrinsic profile. This has been previously assessed by implementing Principal Component Analysis (PCA) of spectra from a large sample of lower-redshift quasars ($z\lesssim2$) where the \lya\ emission is fully observed \citep[e.g.,][]{suzuki_quasar_2006,paris_principal_2011,davies_predicting_2018}. 
We used the PCA model from \citet{bosman_comparison_2021} which we briefly explain here. The PCA training set consists of 4579 quasars from the eBOSS–SDSS DR14 catalog \citep{paris_sloan_2018}.
This method focuses on recovering the \lya\ forest and was repurposed to recover the \lya\ line in 10 quasars at $z\sim6$ from the XQR-30 sample \citep{chen_measuring_2022}. 
The PCA red-side modeling uses the spectrum at wavelengths $\lambda>1290$\AA. The predicted \lya\ profile and bluer continuum is found using a projection matrix relating the red-side PCA coefficients to those in the blue-side of the PCA. 

To apply this updated PCA continuum modeling to the spectrum of P352--15, we first normalize the reduced spectrum from X-shooter observations such that its median flux at $\lambda = 1290\pm2.5$\,\AA\ is unity, following \citet{davies_predicting_2018}. A cubic-spline procedure is applied to automatically and systematically remove absorbers in the spectrum so that the PCA is optimized to model the continuum from the spectrum. Finally, the red-side PCA components are fitted to the spectrum and we obtain the output of the blue-side PCA prediction of the quasar continuum. Figure \ref{spec-pca} shows the PCA model for P352--15 using this method. 
The predicted blue-side spectrum lies well above the observed one, in particular, the predicted \lya\ line is $\sim3$ times weaker than the observed normalized spectrum. This discrepancy at the systemic \lya\ and redward of the profile is then suggestive of a \lya\ damping wing. Note that in the spectrum we also observe a very strong broad \civ\ line from this quasar, which typically strongly correlates with the \lya\ emission \citep[e.g.,][]{greig_lya_2017}.

To calculate the transmitted flux in the proximity zone $R_{\rm p}$, we must know precisely the redshift of the quasar. From this position, we measure the proximity zone up to where flux drops below the 10\% level of the smoothed normalized spectrum. We have a precise systemic redshift of the quasar $z_{\mathrm{sys}}=5.832\pm0.001$ found from \cii\ emission line observations in \citetalias{rojas-ruiz_impact_2021}. The redshift uncertainty is equivalent to 0.06 pMpc. The observed spectrum of P352--15 is normalized by the predicted continuum emission (blue line in Figure \ref{spec-pca}), and is then smoothed with a 20\,\AA\ Boxcar kernel using the Astropy \texttt{convolution}. However, given the caveat of the absorption system presented in the quasar line-of-sight, the normalized spectrum at the position of the quasar is less than half of unity. Following the 10\% level definition of the proximity zone would result in a biased measurement of a very short $R_{\rm p}\sim0.3$\,pMpc (see Appendix \ref{AppB}).

\begin{figure}
\centering
\includegraphics[width=\linewidth]{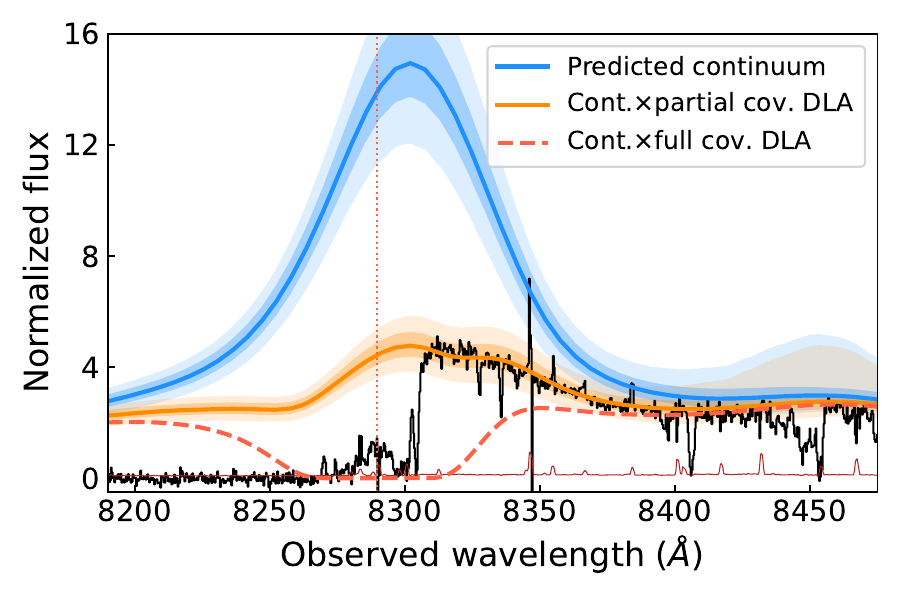}
\caption{P352--15 spectrum (black) and different models predicting the Ly$\alpha$ emission. The model for the predicted unabsorbed spectrum is shown in blue. The DLA system is shown centered at the position of the red dotted vertical line. The continuum fully covered by the DLA is shown with a red dashed line. Given that partial flux is observed, we model the continuum prediction when partially covered by the DLA as shown in orange. Each model has the $\pm1\sigma$ and $\pm2\sigma$ uncertainties in shaded colors.}
\label{ghost}
\end{figure}

The ``shelf" in the smoothed flux in the center of the proximity zone is only a factor of $\sim2$ below 10\% transmission, thus correcting for the absorber would push the value of $R_{\rm p}$ out much further. To address this, we model the spectrum following the scenario where the continuum is partially absorbed by the ghost-DLA system, as shown with the orange solid line and $\pm1\sigma$ and $\pm2\sigma$ uncertainties in Figure \ref{ghost}. This partial DLA model yields a proximity zone of $R_{\rm p}=1.5$\,pMpc. However, this measurement has the caveat that the absorber model is uncertain (i.e. there could be multiple \hi\ components). Another caveat is that, in principle, the absorber could also be covering up the accretion disk, which could also lead to a shorter than average proximity zone. However, we note that at the centered position of the absorber (red dotted vertical line in Figure \ref{ghost}), there is considerable amount of emission. Thus, we continue to estimate an upper limit to the proximity zone size. 

\begin{figure*}[t!]
\centering\includegraphics[width=\linewidth]{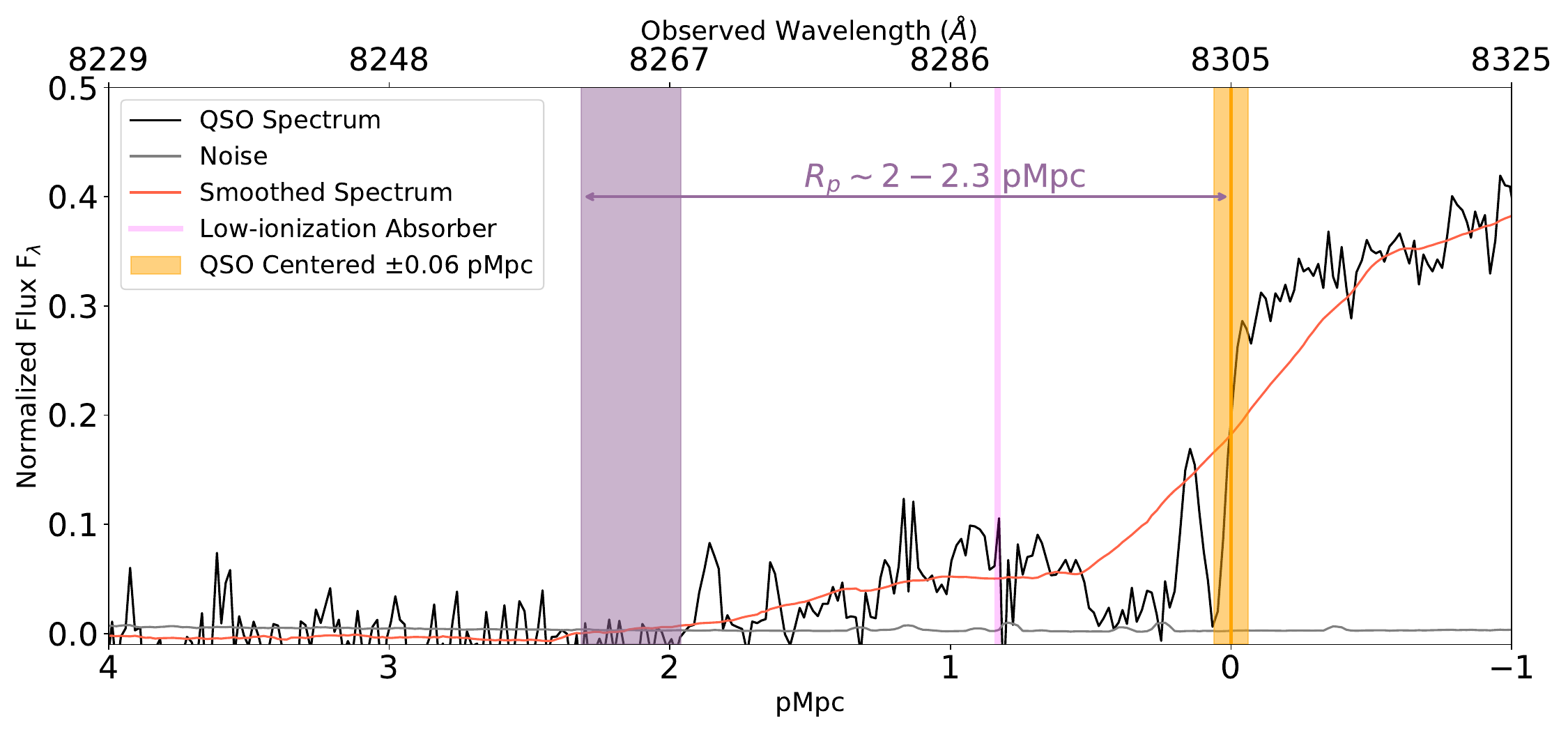}
\caption{Spectrum of P352--15 normalized by the predicted continuum from the PCA analysis (black) and the noise (grey). Given the strong low ionization absorber (centered at the location of the pink vertical line), the normalized flux is less than half at the boundary of the position of the quasar. The flux smoothed with a boxcar kernel of 20\,\AA\ is shown in red. Notice that despite the presence of the strong absorber, significant transmission extends from the position of the quasar (yellow vertical region) up to where the transmission and the smoothed spectrum reach the zero level (shaded in purple). This region corresponds to the measured proximity zone bounds $R_{\rm p}\sim2-2.3$\,pMpc and is the conservative upper limit we use for the calculation of the quasar lifetime $t_{\mathrm{Q}}$.}
\label{Rp}
\end{figure*}

We measure the position where the smoothed spectrum (red line in Figure \ref{Rp}) approaches the zero level that is at $R_{\rm p} = 2.3$\,pMpc. However, by this position a considerable amount of flux is below the noise level (grey line in Figure \ref{Rp}). Thus we set a lower bound to the proximity zone as the position where the transmitted flux (black line in Figure \ref{Rp}) reaches the zero level, that is at $R_{\rm p} = 2$\,pMpc. These values are consistent with the $R_{\rm p}=1.5$\,pMpc from the partially covered DLA model. We use the proximity zone bounds $R_{\rm p} \sim 2-2.3$\,pMpc (purple region in Figure \ref{Rp}) to calculate the quasar lifetime.  

\begin{figure}[t!]
\centering
\includegraphics[width=\linewidth]{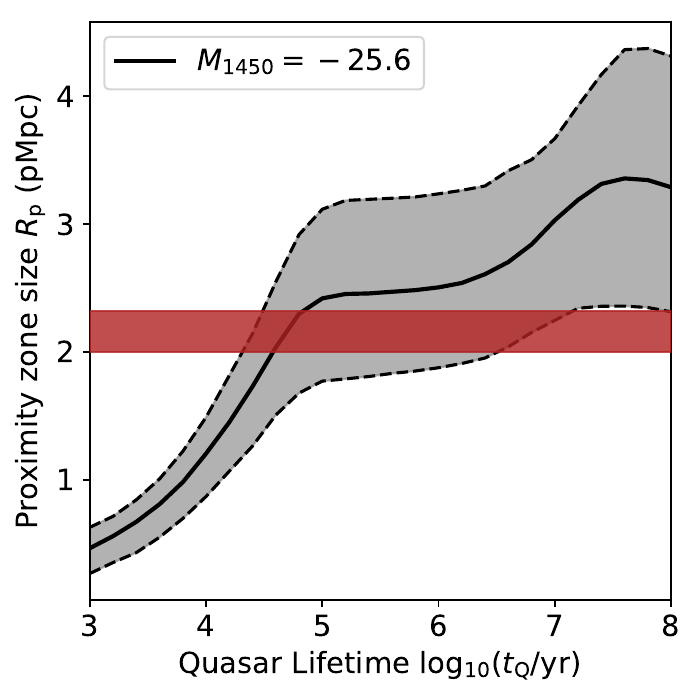}
\caption{Distribution of the evolution of proximity zone sizes with quasar lifetime from radiative transfer simulations for a quasar with $M_{1450}=-25.6$. The black line is the median of the distribution and the shaded grey area represents the 68\% uncertainty. The horizontal red line is located at the proximity zone we measured of $R_{\rm p}\sim2-2.3$\,pMpc. Based on this analysis, the estimated quasar lifetime of P352--15 is ${\rm log}_{10}(t_{\mathrm{Q}}$/yr$) = 4.3-7.1$}
\label{tq}
\end{figure}

\section{Estimating the Quasar Lifetime \label{ql-txt}}
Following the methodology to calculate quasar lifetimes $t_{\mathrm{Q}}$ from proximity zones $R_{\rm p}$ in \citet{eilers_implications_2017}, radiative transfer simulations (RT) for a quasar matching the luminosity and redshift of P352--15 are necessary. The absolute magnitude of the quasar is $M_{1450}=-25.59\pm0.13$ \citep{banados_powerful_2018}. We use RT simulations from \citet{davies_time-dependent_2020} evaluated for $M_{1450}=-25.6$ at $z=6$. Briefly summarized, these simulations solve the photoionization rates for ionized hydrogen and helium species to show the effect of quasar ionizing radiation on the IGM. The initial rapid rise ($t_{\mathrm{Q}} \lesssim 10^5$\,yr) corresponds to hydrogen photoionization that reaches equilibrium by $t_{\mathrm{Q}} \sim 10^6$\,yr before a second increase in transmission at $t_{\mathrm{Q}} \sim 10^7$\,yr as a result of \heii\ reionization heating.

The measured $R_{\rm p} \sim 2-2.3$\,pMpc of P352--15 is small compared to bright quasars with $-27.5 \lesssim M_{1450}\lesssim-26.5$ that have proximity zones $R_{\rm p}= 3 - 7$\,pMpc \citep{eilers_implications_2017}. However, it is close to the mean value for a low-luminous quasar with $M_{1450}=-25.5$ (see \citealt{ishimoto_subaru_2020}). Additionally, a handful of bright quasars similarly show small proximity zones with $R_{\rm p}= 0.3 - 1.9$ pMpc, translating to $t_{\mathrm{Q}}\lesssim 10^4$\,yr \citep{eilers_detecting_2021}. According to the RT simulations, our proximity zone measurement of P352--15 corresponds to a $t_{\mathrm{Q}}$ with 68\% confidence limit as short as ${\rm log}_{10}(t_{\mathrm{Q}}$/yr$) = 4.3$, that can extend even to ${\rm log}_{10}(t_{\mathrm{Q}}$/yr$) = 7.1$ (see Figure \ref{tq}). If our model for the partially-covered ghost-DLA is accurate, the resulting $R_{\rm p}= 1.5$\,pMpc would correspond to a shorter (and more tightly constrained) lifetime of ${\rm log}_{10}(t_{\mathrm{Q}}$/yr$) = 4.2^{+0.4}_{-0.2}$.

To address potential confusion regarding how regions as far as 7 million light-years from the quasar ($R_{\rm p}\sim2.3$\,pMpc) can be affected by activity that began only 40,000\,yr ago $R_{\rm p}\sim2$\,pMpc), it is important to consider the behavior of quasar radiation propagation. While the quasar has become recently active, the proximity zone size can far exceed the light travel time over its observed lifetime. This occurs because the intense ionizing radiation from the quasar propagates vast distances before being depleted. Thus, the photons we observe today originated from the initial wavefront, which traveled significantly farther than the quasar's active age might suggest. Consequently, the observed impact at large distances reflects the early, powerful emission of the quasar, explaining how regions millions of light-years away are influenced within this timeframe.

\section{Discussion and Outlook}\label{discussion-jet}

In this work, we performed a novel experiment aimed at determining the relationship between quasar activity and jet formation. The radio-loud quasar P352--15 at $z\!=\!5.832$ presents the unique opportunity to measure observationally the timescales of these distinct physical processes. For this purpose, we used multi-frequency observations of the quasar from the near-IR to radio regime, and performed independent methodologies. Our findings yielded consistent ages for the latest episode of black hole accretion and the jet in a quasar at the edge of reionization. In this section, we discuss these results and the uncertainties involved in the calculations.

\subsection{Uncertainties in the Jet Age Determination}\label{uncert-jet-age}
The VLA observations of P352--15 presented in this work depict the most detailed characterization in radio frequencies (rest-frame $\sim1-300$\,GHz) of a radio-loud quasar in the epoch of reionization. The jet timescale since launch was calculated using Equation \ref{eq-tdy} and the average synchrotron break frequency from the KP and CI model fits is ($28.76$\,GHz), yielding a jet lifetime of $2,000$\,yr. Using the KP break frequency of $\nu^{\rm break}_{\rm obs} = 39.19 \pm4.6$\,GHz, the lower limit for the dynamical age is $1,700$\,yr. The CI break frequency of $\nu^{\rm break}_{\rm obs} = 18.32\pm1.8$\,GHz gives an upper limit of $2,500$\,yr. Therefore, the jet lifetime is consistent to be $t_{\mathrm{dyn}}\sim10^3$\,yr regardless of the chosen synchrotron cooling model. 

The main caveat on the jet lifetime calculation comes from the accuracy of the magnetic field $B$ measurement. We adopted the average magnetic field value from \citet{momjian_resolving_2018}, calculated using the standard synchrotron minimum energy arguments, as often applied on studies of static radio jets. However, this approach accounts for the internal energy of the jet but neglects the bulk-motion component \citep[e.g.,][]{zdziarski_minimum_2014} and therefore can lead to an imprecise jet power estimate. It is also important to note that it is challenging to calculate the involved energies with only one observation at high-resolution of the jet. Additional multi-frequency observations at high resolution would be necessary to better constrain the magnetic field (Momjian et al., in prep). This would lead to a correction factor of the jet spectral aging of $t_{\rm spec} \sim (\frac{B}{3.5 \mathrm{mG}})^{-1.5}$ according to Equation \ref{tspec-eqn}, and for the dynamical age a factor of $t_{\rm dyn} \sim (\frac{B}{3.5 \mathrm{mG}})^{2}$ following Equation \ref{eq-tdy}. 

Finally, we note that there is a remaining observational gap of P352--15 between $\sim 42$\,GHz and 100\,GHz. Supplemental observations could allow for a clearer distinction between the KP and CI models, leading to a revised calculation for spectral and dynamical ages of the jet. The only telescope that could help partially fill in the gap is ALMA but only from $\sim70$\,GHz, and the sensitivity at this frequency is challenging for the required observations. However, our analysis thus far shows that the observed emission at 100\,GHz as measured with NOEMA in \citetalias{rojas-ruiz_impact_2021} is mostly of non-thermal nature instead of cold dust. In order to properly study this interconnection between cold dust and synchrotron emission in P352--15, it is necessary to more accurately model the cold dust with an MBB model. This can be achieved with further sub-millimeter observations filling the gap between 100\,GHz and 290\,GHz. ALMA, NOEMA, or The Submillimeter Array (SMA) interferometers are able to reach the required sensitivity for these observations. Importantly, obtaining robust 5$\sigma$ limit observations at 450\,GHz would help constrain the MBB peak and thus the cold dust temperature of the host galaxy of P352--15.

\subsection{Superluminal Motion and Boosting of the Radio-jet}
\citet{momjian_resolving_2018} reveal a one-sided jet that extends up to a distance $d_{\rm jet} \approx1.62$\,kpc. This is the projected distance of the jet observed from an angle $\theta$ to the line-of-sight of the observer. By measuring this angle, it is possible to constrain the superluminal motion and Doppler boosting effect of the jet. The angle can be calculated by using the following relationship:

\begin{equation}\label{betaT}
\beta_T=\frac{\beta_{\rm jet}\ \mathrm{sin}\ \theta}{ 1 - \beta_{\rm jet} \ \mathrm{cos}\ \theta}
\end{equation}

Here, $\beta_T$ relates to the apparent transverse velocity which is found using the observed distance of the jet hot spots and the calculated dynamical age such that $\beta_T c= d_{\rm jet}/t_{\mathrm{dyn}}$.  We calculate a superluminal motion of the jet of P352--15 with $\beta_T = 2.6$. The jet bulk velocity $\beta_{\rm jet}$ is taken from Equation \ref{Gammajet}. Solving for the viewing angle in Equation \ref{betaT}, we obtain $\theta \approx 25^\circ$. Note that blazars have $\theta \lesssim 20^\circ$, Fanaroff-Riley radio galaxies have $\theta \sim 90^\circ$, and quasars have angles in-between.

We can measure the Doppler boosting effect $\delta$ using the calculated viewing angle $\theta$ and the previously calculated values of $\Gamma_{\rm jet}=2.35$ and $\beta_{\rm jet}=0.90$ from Equation \ref{Gammajet}, such that:

\begin{equation}\label{boost}
\delta=\frac{1}{\Gamma_{\rm jet} (1 - \beta_{\rm jet} \ \mathrm{cos}\ \theta)}
\end{equation}

We obtain a Doppler boosting of $\delta \approx 2.7$. As the viewing angle of the jet with respect to the line-of-sight decreases, the boosting or beaming is stronger. This effect often prevents the radio counterjet from being observed as its apparent luminosity is dimmer compared to the intrinsic one, as in the case of P352--15. Similarly, the brightness of the observed jet increases as it is boosted by a factor $\delta^n$, where $n$ has a value between 2 and 3 depending on the jet geometry and spectral index \citep{kellermann_radio_1988,cohen_relativistic_2007,kellermann_doppler_2007}. Accordingly, the radio emission would be boosted by a factor of 7--20 for the case of P352--15. Further multi-epoch observations at high angular resolution with the VLBA are necessary to get more accurate motions from apparent luminosities of the hot spots, and therefore the corresponding intrinsic brightness.

\subsection{Caveats on the Measured Quasar Lifetime}
The calculated lifetime for the quasar activity for P352--15 is achieved by estimating its proximity zone, following the method developed in \citet{eilers_implications_2017}. When analyzing the quasar spectrum, we encountered a strong low ionization absorber that could possibly truncate the proximity zone. However, we see substantial flux transmission even at the central position of the absorber. Thus we could estimate a limit on the proximity zone corresponding to a quasar lifetime $t_{\mathrm{Q}}\sim10^4$\,yr. The rare ghost-DLA system is very interesting, and the fact that another highly radio-loud quasar at $z\sim2.1$ \citep{gupta_h_2022} shows a similar feature is suggestive of a connection between the two. Given that both sources are separated by billions of years of evolution, this correlation suggests that we are gaining a deeper understanding of the physical processes occurring in the BLR of AGNs.

The calculated timescales for the jet and quasar lifetimes are found to be consistent despite their significant systematic uncertainties and wildly different physical phenomena.
This could indicate a connection between accretion disk activity and jet launch. The implied activity timescale of $\lesssim 10^4$\,yr is too small to grow the SMBH even at `super-Eddington' ratio regimes \citep[see also ][]{eilers_first_2018,eilers_detecting_2020,eilers_generative_2022}. Analysis on the quasar's \civ\ emission result in the black hole accreting at an Eddington ratio of $\lambda_{\rm Edd} = 0.28^{+0.47}_{-0.17}$ (Xie et al. in prep.), which is a typical value among high-redshift quasars \citep[e.g.,][]{farina_x-shooteralma_2022}. This might suggest that the quasar is flickering \citep[e.g.,][]{davies_time-dependent_2020,satyavolu_need_2023}. This hypothesis could be tested by looking for the presence of jet emission at larger kpc-scales that would hint at a previous black hole-jet activity. In fact, \citet{connor_enhanced_2021} suggested the presence of a possible X-ray jet at projected $\sim50$\,kpc from the quasar center. If this jet has a radio counterpart, then this would correspond to a jet launch occurring before the timescale we calculated in this work where the extension of the jet is just $\sim1.6$\,kpc.  

Alternatively, a potential scenario from our results is a mismatch in the jet and quasar lifetimes resulting from UV-obscured accretion. Indeed, $z \gtrsim 7$ quasars appear to favor long periods of SMBH growth with substantial dust obscuration \citep[e.g.,][]{davies_evidence_2019} that does not affect radio-jet activity. In fact, the radio jet can contribute to the dispersal of the obscuring dust, leading to a larger jet than quasar lifetime.

In summary, the quasar P352--15 provides a unique opportunity to study the connection between the launching of jets and accretion onto the black hole. It provides the possibility to measure at the same time the break in its synchrotron spectrum to determine the age of the radio jet and the proximity zone to estimate the most recent black hole accretion timescale. Most importantly, this experiment could be repeated for the other radio-loud quasars at the epoch of reionization (e.g. \citealt{willott_canadafrance_2010,banados_constraining_2015,banados_pan-starrs1_2023,banados_blazar_2024, belladitta_first_2020,gloudemans_discovery_2022, ighina_radio_2021,ighina_multi-wavelength_2024,wolf_srgerosita_2024}). Indeed, this study establishes a valuable precedent for exploring radio quasars at $z\gtrsim 6$, paving the way for future studies that will further elucidate the relationship between jet activity and black hole accretion.
\acknowledgements
 We thank Christian Fendt for the fruitful discussion on jet physics and the interpretation of our results. We thank Zhang-Liang Xie \added{and Frederick Hamman} for helpful conversations. S.R.R. acknowledges financial support from the International Max Planck Research School for Astronomy and Cosmic Physics at the University of Heidelberg (IMPRS--HD). C.M. acknowledges support from the ANID BASAL project FB210003. The National Radio Astronomy Observatory is a facility of the National Science Foundation operated under cooperative agreement by Associated Universities, Inc.  This paper makes use of VLA data from program 21A-307. Based on observations collected at the European Southern Observatory under ESO programme 0102.A-0931.

\facility{VLA, VLT}
\software{Astropy \citep{the_astropy_collaboration_astropy_2013}, CASA \citep{mcmullin_casa_2007},
          linetools         \citep{prochaska_linetoolslinetools_2017},
          Matplotlib \citep{hunter_matplotlib_2007},
          Numpy \citep{harris_array_2020},
          PypeIt
          \citep{prochaska_pypeit_2020},
          SciPy
          \citep{virtanen_scipy_2020}.}
          
\bibliographystyle{yahapj}
\bibliography{references.bib}      

\appendix
\section{Imaging from the VLA Observations}\label{Appendix-vla1}
\begin{figure*}[hbpt!]
\centering

\includegraphics[width=1\linewidth]{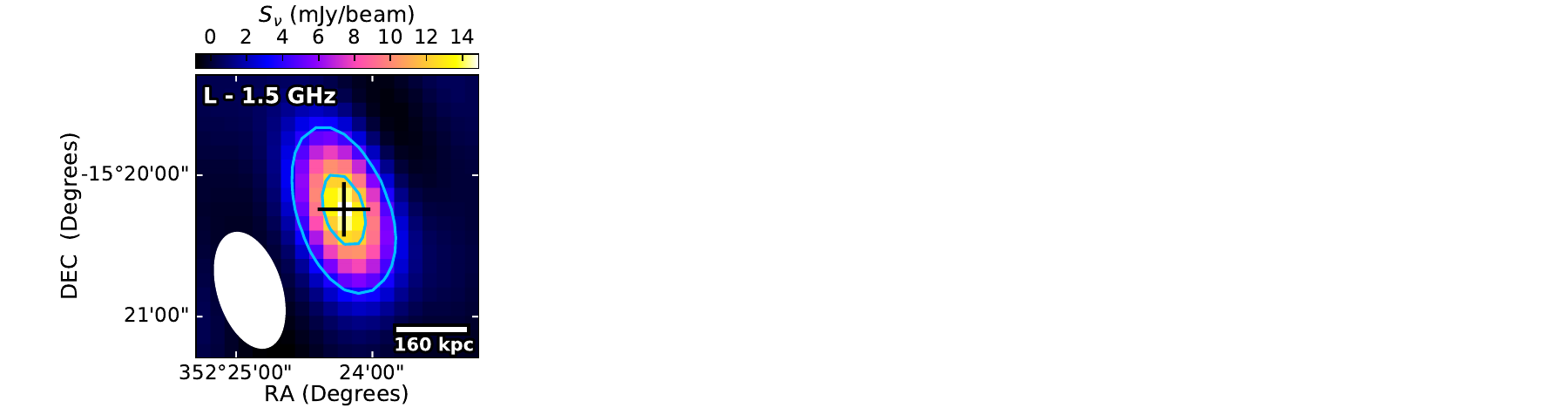}

\includegraphics[width=1\linewidth]{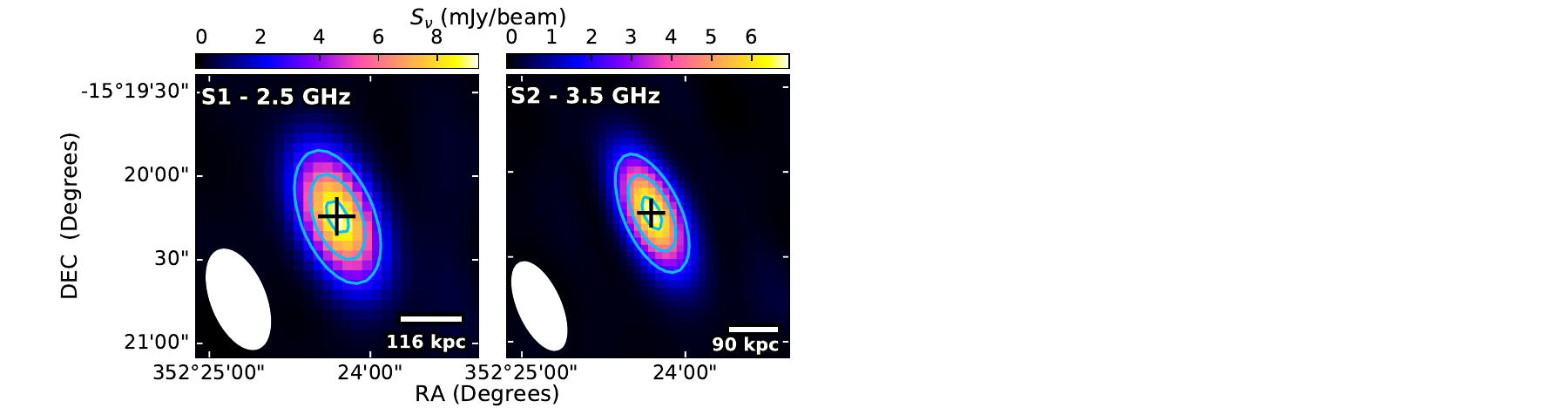}

\includegraphics[width=1\linewidth]{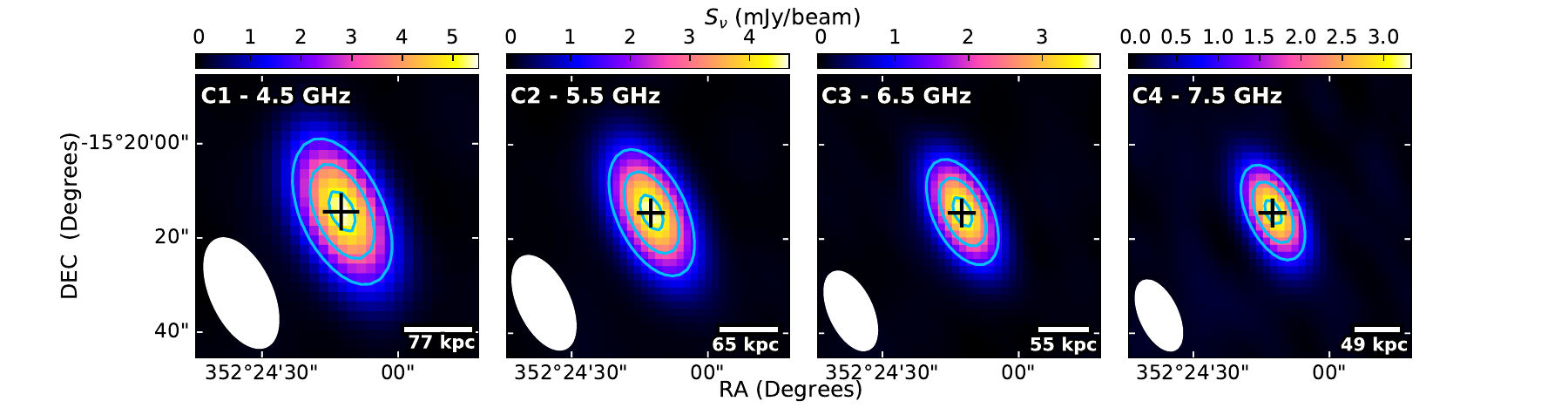}

\caption[Reduced VLA images in L\,S\,C bands]{Reduced images from the L-band and S- and C-band chunks. In all stamps, the color map indicates the flux density $S_\nu$ of the synchrotron emission. The central cross represents the optical position of P352--15 as seen in \citet{momjian_resolving_2018}. Each stamp has the central frequency of the observation at the top and the beam size at the bottom left corner. Note that for visualization purposes, each set of sub-bands has been zoomed in (see the physical scale at the bottom right). The contours are set as follows: L-band [$-1, 1, 3]\times\sigma$, S and C-bands [$-1, 6, 12, 18]\times\sigma$. Refer to Table \ref{vla-tab-data} for the corresponding $\sigma$ errors and S/N of each sub-band. While the negative contours are present, they are not visible in the zoomed-in portion of the image.}

\label{LSCX}
\end{figure*}

\begin{figure*}[hbpt!]
\centering
\includegraphics[width=1\linewidth]{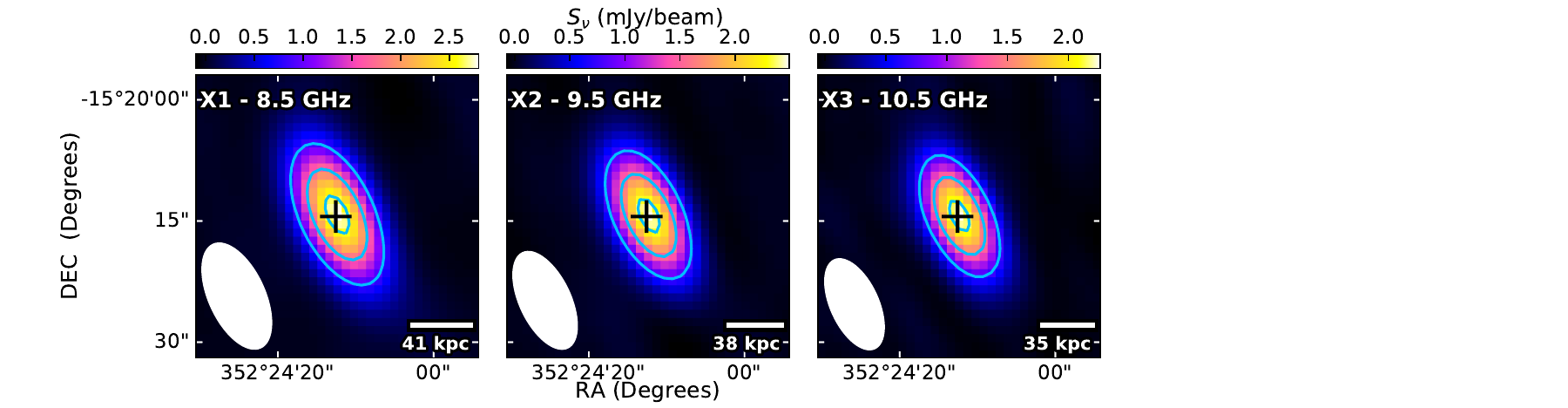}

\includegraphics[width=1\linewidth]{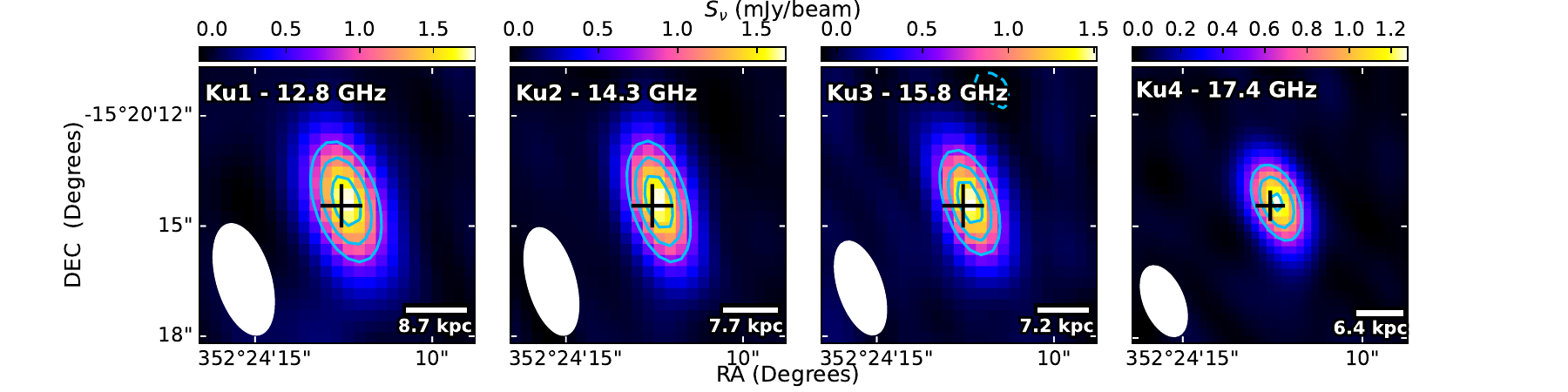}

\includegraphics[width=1\linewidth]{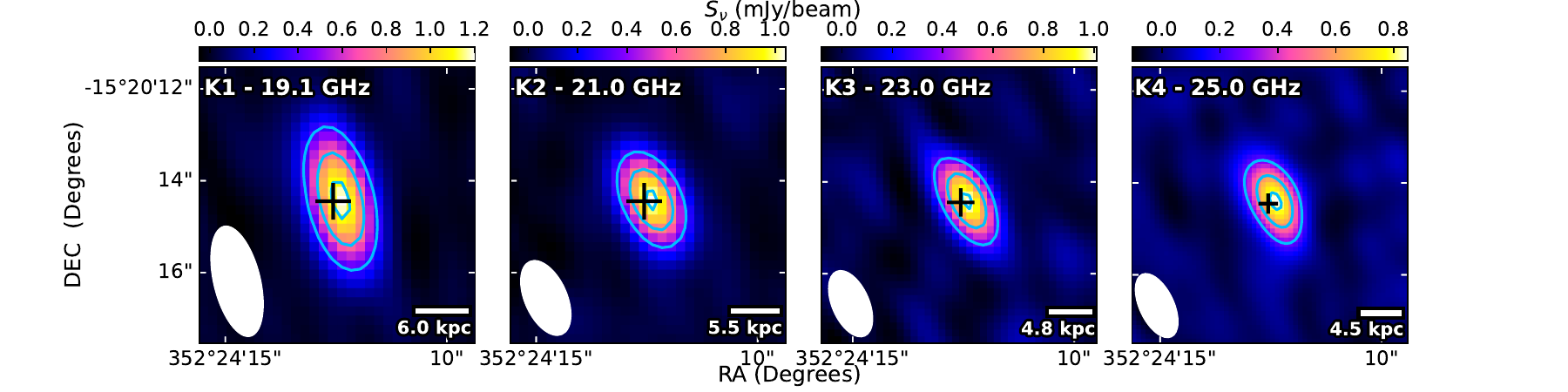}

\includegraphics[width=1\linewidth]{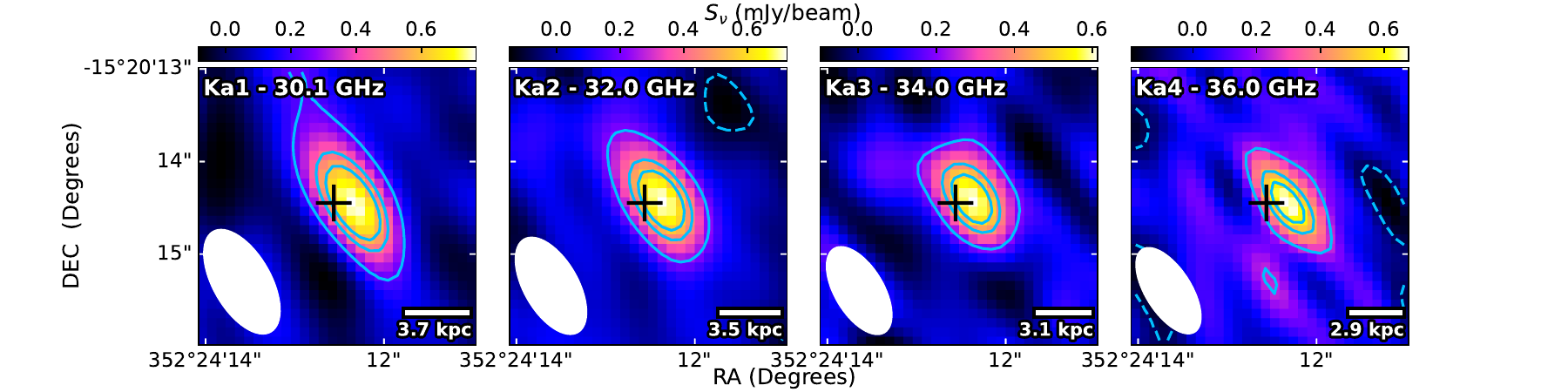}
\includegraphics[width=1\linewidth]{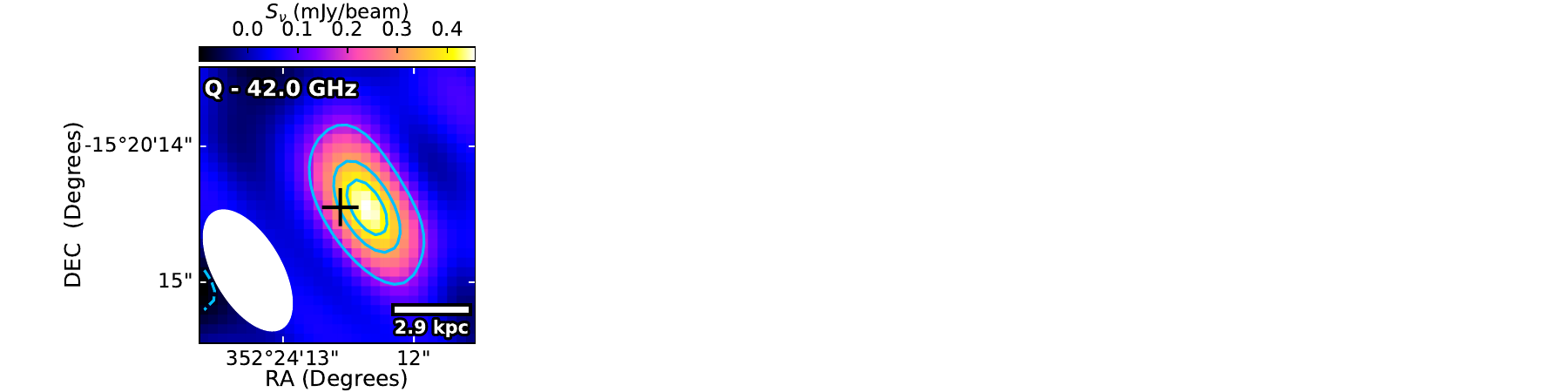}
\caption[Reduced VLA images in X\,Ku\,K\,Ka\,Q bands]{Similar to Figure \ref{LSCX} but for the X, Ku, K, Ka, and Q observations. The contours are set as follows: X-bands  [$-1, 6, 12, 18]\times\sigma$, Ku-bands [$-1, 8, 12, 16]\times\sigma$, K-band [$-1, 3, 6, 9]\times\sigma$, Ka- and Q-bands [$-1, 2, 4, 5]\times\sigma$. Negative contours are shown in dashed cyan lines, where visible.}
\label{UKAQ}
\end{figure*}

This section presents the reduced images from our quasi-simultaneous multi-frequency VLA observations. The data span all VLA frequency bands from L-band to Q-band. Figures \ref{LSCX} and \ref{UKAQ} display images for each frequency band, organized by row, corresponding to the selected frequency chunks as outlined in \S \ref{vla-obs}. Flux measurements were obtained by fitting a 2D Gaussian to these images. These measurements were then used to model the combined cold dust and synchrotron emission from the P352--15 quasar, as illustrated in Figure \ref{sync-models}.

\newpage

\section{Proximity zone}\label{AppB}
\begin{figure*}[hbpt!]
\centering
\includegraphics[width=1\linewidth]{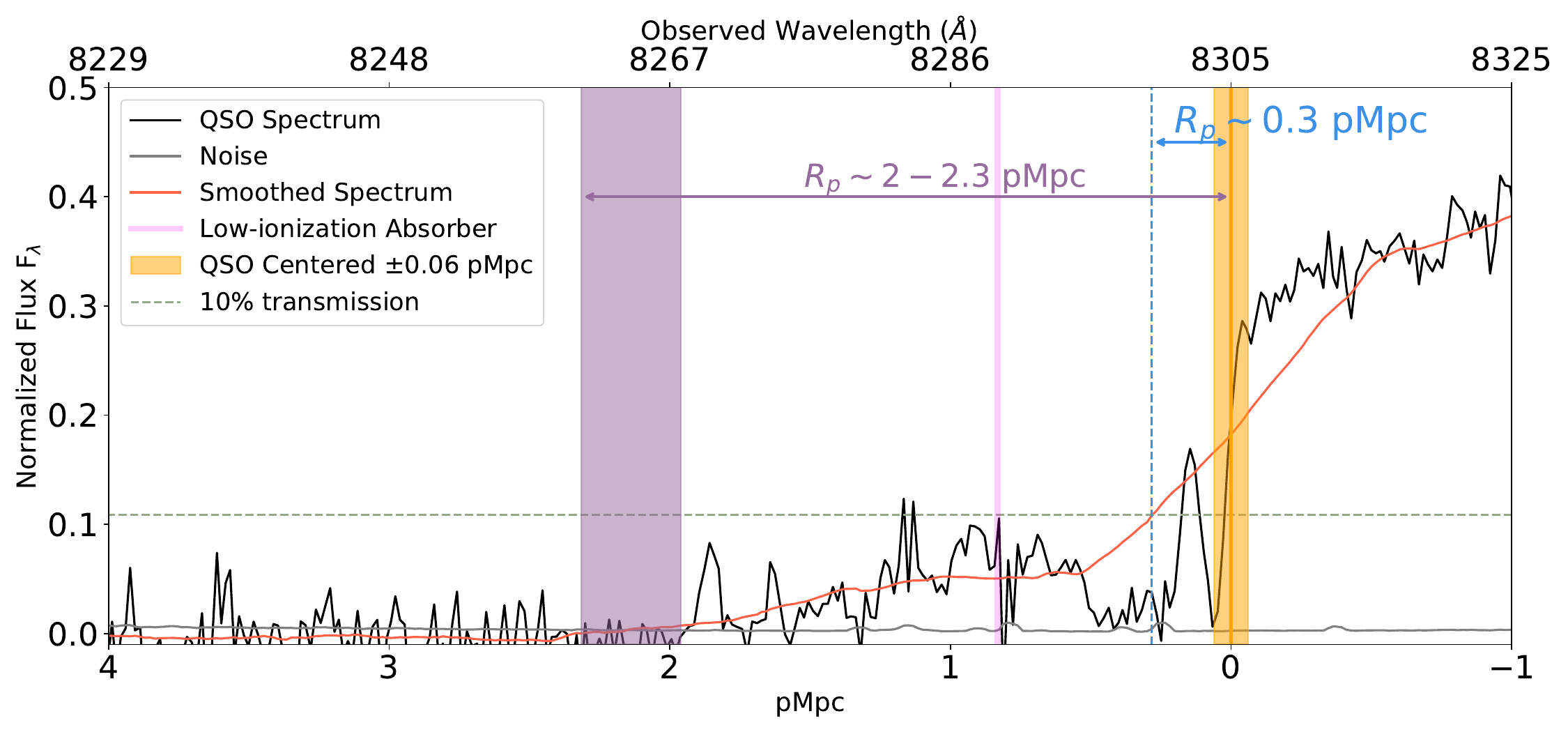}
\caption{This figure shows the normalized spectrum of P352--15 (black) and the noise (grey). The flux smoothed with a boxcar kernel of 20\,\AA\ is shown in red. The position of the quasar is shown with the yellow region and the horizontal dashed line indicates 10\% of the normalized spectrum. The vertical dashed line in blue shows the estimate for the proximity zone $R_{\rm p}\sim0.3$\,pMpc following the traditional procedure to measure proximity zones. The strong low-ionization absorber is centered at the location of the magenta vertical line. The vertical purple region shows the proximity zone measurement when the transmitted and smoothed flux reach the zero level, $R_{\rm p}\sim2-2.3$\,pMpc.}
\label{Rp_10p}
\end{figure*}

This section shows in detail the proximity zone measurements given the traditional definition and the method we adopted for the paper. Figure \ref{Rp_10p} illustrates the proximity zone $R_{\rm p}$ measurements given the accepted definition where $R_{\rm p}$ is measured from the position of the quasar to where the smoothed flux drops below the 10\% level. However, given the caveat of the absorption system presented in the quasar line-of-sight, the resulting proximity zone measurement is biased to a very short $R_{\rm p}\sim0.3$\,pMpc. Given that significant transmission extends even further out, we chose to measure the $R_{\rm p}$ from a lower bound where the transmitted flux (black line) reaches zero to an upper bound where the smoothed spectrum (red line) reach the zero level. The conservative upper limit for the proximity zone is then $R_{\rm p}\sim2-2.3$\,pMpc that is used to calculate the quasar lifetime $t_{\mathrm{Q}}$ in \S\ref{ql-txt} and for the interpretation of our results.

\end{document}

%% file: definitions.tex
\definecolor{red}{rgb}{1,0,0}
\definecolor{orange}{RGB}{204, 85, 0}
\definecolor{blue}{HTML}{4169e1}
\definecolor{ltred}{RGB}{245,167,162}
\definecolor{ltblue}{RGB}{206,211,242}

\newcommand{\lya}{Ly$\alpha$}

\newcommand{\hi}{H\,{\sc i}}
\newcommand{\hii}{H\,{\sc ii}}

\newcommand{\heii}{He\,{\sc ii}}

\newcommand{\siiv}{Si\,{\sc iv}}

\newcommand{\civ}{C\,{\sc iv}}
\newcommand{\nv}{N\,{\sc v}}
\newcommand{\cii}{[C\,{\sc ii}]}
\newcommand{\ciiS}{C\,{\sc ii}}

\newcommand{\oi}{O\,{\sc i}}

\newcommand{\siii}{Si\,{\sc ii}}

\newcommand{\alii}{Al\,{\sc ii}}

\newcommand{\Mdust}{\textit{M}$_{\mathrm{Dust}}$ }

\newcommand{\Msun}{$M_\odot$}
\newcommand{\angstrom}{\text{\normalfont\AA}}

\defcitealias{rojas-ruiz_impact_2021}{RR21}